\pdfoutput=1
\documentclass[numberedappendix]{emulateapj}
\usepackage[breaklinks,colorlinks,citecolor=blue,linkcolor=magenta]{hyperref}
\usepackage{natbib}
\usepackage{psfig}
\usepackage{amsmath}

\newcommand{\pdif}[2]{\ensuremath{ \frac{\partial #1}{\partial #2}}}

\def\bfnabla{{\mbox{\boldmath $\nabla$}}}

\newcommand\bn{{\mbox{\boldmath $n$}}}

\newcommand\Crat{{\mathbb{C}}}
\newcommand\Prat{{\mathbb{P}}}
\newcommand\Rrat{{\mathbb{R}}}

\def\<{\,\langle\langle}
\def\>{\,\rangle\rangle}

\shorttitle{Dusty Cloud Acceleration}
\shortauthors{Zhang et al.}

\begin{document}


\title{Dusty Cloud Acceleration by Radiation Pressure in Rapidly Star-Forming Galaxies}


\author
{Dong Zhang\altaffilmark{1}, 
Shane W. Davis\altaffilmark{1},
Yan-Fei Jiang\altaffilmark{2},
and James M. Stone\altaffilmark{3}}

\affil{$^{1}$Department of Astronomy, University of Virginia, Charlottesville, VA 22904, USA}
\affil{$^{2}$Kavli Institute for Theoretical Physics, University of California, Santa Barbara, CA 93106, USA}
\affil{$^{3}$Department of Astrophysical Sciences, Princeton University, Princeton, NJ 08544, USA}
\email{dz7g@virginia.edu}







\begin{abstract}

We perform two-dimensional and three-dimensional radiation hydrodynamic simulations to study cold clouds accelerated by radiation pressure on dust in the environment of rapidly star-forming galaxies dominated by infrared flux. We utilize the reduced speed of light approximation to solve the frequency-averaged, time-dependent radiative transfer equation.  We find that radiation pressure is capable of accelerating the clouds to hundreds of kilometers per second while remaining dense and cold, consistent with observations. We compare these results to simulations where acceleration is provided by entrainment in a hot wind, where the momentum injection of the hot flow is comparable to the momentum in the radiation field. We find that the survival time of the cloud accelerated by the radiation field is significantly longer than that of a cloud entrained in a hot outflow. We show that the dynamics of the irradiated cloud depends on the initial optical depth, temperature of the cloud, and the intensity of the flux. Additionally, gas pressure from the background may limit cloud acceleration if the density ratio between the cloud and background is $\lesssim 10^{2}$. In general, a 10 pc-scale optically thin cloud forms a pancake structure elongated perpendicular to the direction of motion, while optically thick clouds form a filamentary structure elongated parallel to the direction of motion. The details of accelerated cloud morphology and geometry can also be affected by other factors, such as the cloud lengthscale, the reduced speed of light approximation, spatial resolution, initial cloud structure, and the dimensionality of the run, but these have relatively little affect on the cloud velocity or survival time.
\end{abstract}



\section{Introduction}

Galactic winds are ubiquitous in rapidly star-forming galaxies and starbursts. They are crucial to the formation and evolution of galaxies, shaping the galaxy luminosity function (\citealt{Benson03}; \citealt{Bower12}; \citealt{PS13}), affecting the chemical evolution of galaxies (\citealt{Erb08}; \citealt{PS11}; \citealt{Barai13}; \citealt{Makiya14}), determing the mass-metallicity relation (\citealt{Tremonti04}; \citealt{FD08}), and regulating star and galaxy formation over cosmic time and polluting the intergalactic medium with metals (\citealt{Aguirre01a}; \citealt{Aguirre01b}; \citealt{Oppenheimer10}). Many mechanisms have been proposed for driving galactic winds, including hot flow heated and driven by supernova explosions (\citealt{Larson74}; \citealt{DS86}; \citealt{CC85}; \citealt{SH09}), radiation pressure by absorption in spectral lines or the continuum absorption and scattering of starlight on dust grains (\citealt{Proga98}; \citealt{Murray05}; \citealt{Murray10}; \citealt{Hopkins12}), cosmic rays (\citealt{Ipavich75}; \citealt{Socrates08}; \citealt{Uhlig12}), and photoevaporation heating by hot stars (\citealt{BL99}; \citealt{Shapiro04}; \citealt{Iliev11}). 

Galactic winds have multi-phases over a broad range of temperature. Multiwavelength observations from radio to X-ray have been carried out to probe all types of gas and dust in winds (see \citealt{Veilleux05}, and references therein). For example, soft and hard X-rays, which are observed by the \textit{Chandra X-ray observatory} and \textit{XMM-Newton observatory}, have been used to constrain hot gas outflows (e.g., \citealt{SH07,SH09}; \citealt{Zhang14}; \citealt{Chisholm17}). Also, emission lines such as H$\alpha$, N II, O II, O III, and absorption lines such as Na I, K I, Mg II from near-IR and optical are observed to trace cold and warm gas in outflows, including molecular (e.g., \citealt{Sakamoto99}; \citealt{Veilleux09}; \citealt{Cicone14}), neutral atomic (e.g., \citealt{Heckman00}; \citealt{Rupke02, Rupke05a, Rupke05b, Rupke05c}; \citealt{Martin05}; \citealt{Kornei13}), and ionized gas (\citealt{SB98}; \citealt{Martin98}; \citealt{Cooper09}).

The formation and acceleration of cold clouds in winds are still unknown. It is commonly believed that cold clouds are advected into winds. The prevailing picutre is that these clouds are embedded in a hot wind, and driven out of the host galaxy by ram pressure of the hot wind (e.g., \citealt{Murray07}; \citealt{SH09}). However, clouds may be shredded and destroyed by hydrodynamic instabilities long before they reach the velocities reqruied by observations (\citealt{Klein94}; \citealt{Poludnenko02}; \citealt{SB15}; \citealt{BS16}). It is uncertain whether cloud magnetization or fragmentation will significantly suppress the cloud disruption and push clouds to be co-moving with the hot wind (\citealt{Cooper09}; \citealt{McCourt15, McCourt16}). Recently, \cite{Thompson16} suggested that although the initial cold clouds might be rapidly destroyed in the hot winds on small scales, the hot flow may cool radiatively on larger scales, forming an extended region of cool gas (see also \citealt{Scannapieco17}). Then cold clouds can be self-induced by thermal instability and re-born again in the halo of host galaxy.

On the other hand, the radiation pressure of starlight on dust is also suggested as another mechanism to accelerate cold clouds in star-forming galaxies (\citealt{Murray05}; \citealt{Murray11}). The evolution of dusty shells in the radiation-dominated regime has been well studied, both semi-analytically (\citealt{Thompson15}), and numerically (\citealt{KT12, KT13}; \citealt{Davis14}; \citealt{Tsang15}; \citealt{Rosdahl15}; \citealt{ZD17}). The dynamics of cloud acceleration by radiation pressure on dust is different from the dynamics of a dusty shell, because clouds only cover a small fraction of the solid angle from the host galaxy, and the cloud expansion behaves differently from a shell. It has been proposed that cold clouds may be ejected above the disk of the host galaxy by radiation from a single massive star cluster before any supernova explode, then radiation pressure from starlight in the disk may drive clouds to a distant of $\sim100$ kpc (\citealt{Murray05}). If the cloud is not pressure confined by its surrounding medium, it will expand at its internal sound speed, and reach very high velocities (\citealt{Murray11}; \citealt{Thompson15}). However, all of the previous semi-analytic works are over-simplified in their assumptions or calculations. For example, the cloud is assumed to be a sphere, and turbulence and shredding are neglected in the cloud. The fully multi-dimensional radiation hydrodynamic simulations are needed to model the complicated properties and behaviors of the cloud during its acceleration and destruction. 

A variety of algorithms for solving radiation hydrodynamics equations have been developed in the literature, including the traditional flux-limited diffusion method (\citealt{Levermore81}; \citealt{Krumholz07}; \citealt{Turner01}; \citealt{Zhang11}; \citealt{KT12, KT13}), the M1 closure method (\citealt{Gonzales07}; \citealt{Skinner13,Skinner15}; \citealt{McKinney14}; \citealt{Rosdahl15}), the variable Eddington tensor (VET) method (\citealt{Stone92}; \citealt{Sekora10}; \citealt{Davis12}; \citealt{Jiang12}), and the Monte Carlo (MC) method (\citealt{Whitney11,Tsang15,Roth15, Bisbas15}). A comprehensive comparison of radiation hydrodynamic codes has not been carried out, but disagreements have been found for some specific problems. A code comparison similar to that carried out in \cite{Bisbas15} for ionizing radiation would be benefited. 


For the problem of the radiation-pressure-dominated flow and optically-thick dusty gas, \cite{KT12} studied the acceleration of dusty shells in a constant infrared radiation flux with a constant gravitational acceleration using the FLD method, and \cite{KT13} studied the interaction between a constant infrared flux and dusty gas but without gravity. They found that radiative Rayleigh-Taylor instability limits momentum transfer from radiation to gas to $\sim L/c$, where $L$ is the radiation luminosity. \cite{Rosdahl15} found similar results using the M1 closure method . \cite{Davis14} revisited the problems in \cite{KT12} using the VET method which directly solves the time-independent radiative transfer equation based on the short characteristic method, and found a stronger momentum coupling between radiation and gas compared to the results in \cite{KT12} in the systems with the initial infrared optical depth $\tau_{\rm IR(initial)}\sim 1-10$. 
Similar results were found in \cite{Tsang15} who used the MC method. Moreover, in \cite{ZD17} we revisited the problems in \cite{KT13} using the same VET method as in \cite{Davis14} and also found more efficient momentum coupling between radiation and gas compared to the FLD results: the momentum transfer from radiation to gas is $\sim (1+\epsilon) (L/c)$ with a boost factor $\epsilon\sim 1.6-23$ for $\tau_{\rm IR(initial)}\sim 1-10$. Since they directly solve the radiation transfer equation, the VET and MC methods are expected to give more accurate results than the FLD and M1 methods for dusty gas with low to moderate optical depth.

However, both the VET and the MC method have their own disadvantages. The intrinsic noise associated with Monte Carlo makes this method very expensive to simulation the dusty gas. The VET method uses the implicit differencing method which needs to invert large matrices every time step, therefore the VET method is also expensive. As an alternative to the VET method, an algorithm using explicit differencing of spatial operators to solve the time-dependent radiative transfer equation has been developed by \cite{Jiang14}. This radiation module was implemented in the MHD code \textsc{athena} (\citealt{Stone08}). Recently, a new radiation module (\citealt{Jiang16, Jiang17}) which use a similar algorithm as in \cite{Jiang14}, has been implemented in a new MHD code \textsc{athena++}, which is the upgraded version of \textsc{athena} (\citealt{White16}; Stone et al. 2017 in preparation). Radiation hydrodynamic simulations have explored the evolution of irradiated clouds in the active galactic nucleus (AGN) context (\citealt{Schartmann11}; \citealt{Proga14}; \citealt{Namekata14}; \citealt{Proga15}; \citealt{Waters16, Waters17}). In particular, \cite{Proga14} investigated the effects of irradiation with various absorption and scattering opacities on the evolution of a cloud in the AGN environment, using the VET method with \textsc{athena}. So far, no work has been done to simulate the cloud acceleration and evolution by radiation pressure on dust in a radiation-dominated regime. This paper is the first work to explore this question using the new radiation MHD \textsc{athena++} code. 

This paper is organized as follows. In Section \ref{section_setup} we describe the radiation hydrodynamic equations, numerical methods, the reduced speed of light approximation, and the simulation setup. The parameters for our simulations are summarized in Table \ref{tab_parameter}. We show the simulation results in Section \ref{section_results}, with different cloud characteristic lengthscales, background temperature, reduction factor for the speed of light, spatial resolution, and simulation dimensionality. In Section \ref{section_discussion} we compare the cloud survival time in a radiation field to that in a hot flow, and discuss some secondary factors which may affect the cloud evolution. We also discuss the implications of the simulation results in Section \ref{section_discussion}. Conclusions are provided in Section \ref{section_conclusions}.

\section{Methods and Initial Conditions}\label{section_setup}

\subsection{Equations}

We solve the equations of mass, momentum, energy with radiation
\begin{eqnarray}
&&\pdif{\rho}{t} + \mathbf{\nabla} \cdot \left(\rho \mathbf{v} \right) = 0\label{hydroequation1}, \\
&&\pdif{\left(\rho\mathbf{v}\right)}{t} + \mathbf{\nabla} \cdot \left( \rho\mathbf{v} 
\mathbf{v} + {\sf P}\right) =  - \mathbf{S}_r(\mathbf{P})\label{hydroequation2}, \\
&&\pdif{E}{t} + \mathbf{\nabla} \cdot \left(E \mathbf{v} + {\sf P} \cdot \mathbf{v}\right) = -c S_r(E)\label{hydroequation3}. 
\end{eqnarray}
Here $\rho$, $\mathbf{v}$, $\mathbf{g}$ are the gas density, fluid velocity and the gravitational acceleration, $E=p/(\gamma-1)+\rho v^{2}/2$ is the total fluid energy density, with $p$ being the gas pressure and $\gamma=5/3$ being the gas adiabatic index. Also $\mathbf{S}_r(\mathbf{P})$ and $S_r(E)$ are the radiation source terms, which are discussed below.
The time-dependent radiative transfer (hereafter RT) equation is
\begin{equation}
 \frac{\partial I}{\partial t}+c\bn\cdot\bfnabla I= S(I,\bn).\label{RTequation}
\end{equation}
The RT equation is similar as in \cite{Jiang16}. The basic algorithm for solving the RT equation was first described by \cite{Jiang14}, then modified by \cite{Jiang16} (see also \citealt{Jiang17}). The source terms $S(I,\bn)$, $\mathbf{S}_r(\mathbf{P})$ and $S_r(E)$ are no longer expanded to the first order of $\mathcal{O}\left(v/c\right)$ as in \cite{Jiang14}. Instead, the specific intensity $I(\bn)$ with angle $\bn$ in the lab frame is first transformed to the co-moving frame intensity $I_0(\bn_0)$ via Lorentz transformation, where $\bn_0$ is the angle in the co-moving frame. The source term $S$ in the co-moving frame is
\begin{eqnarray}
 S(I_{0},\bn_{0})&=&c\rho \kappa_{aR}\left(\frac{a_rT^4}{4\pi}-I_0\right)+c\rho \kappa_ s (J_0-I_0)\nonumber\\
&& + c\rho (\kappa_{aP}-\kappa_{aR})\left(\frac{a_r T^{4}}{4\pi}-J_0\right),\label{SourceTerm1}
\end{eqnarray}
where $\kappa_{aR}$ and $\kappa_{aP}$ are the Rosseland and Planck mean absorption opacities, $\kappa_s$ is the scattering opacity respectively, and $J_0 = \int I_0 (\bn_0)d\Omega_0$ is the angular quadrature of the specific intensity in the co-moving frame. The radiation momentum and energy source terms $\mathbf{S}_r(\mathbf{P})$ and $S_r(E)$ in the co-moving frame are given by
\begin{equation}
\mathbf{S}_r(\mathbf{P})  = -\frac{\rho(\kappa_s+\kappa_{aR})}{c}\mathbf{F}_r,
\end{equation}
and
\begin{equation}
S_r(E)  = \rho \kappa_{aP}(a_rT^4-E_r)
\end{equation}
respectively, where $\mathbf{F}_r$ is the radiation flux, and $E_r$ is the radiation energy. $I_0(\bn_0)$ is updated implicitly in the co-moving frame, and transformed back to the lab frame again via Lorentz transformation. The above radiation hydrodynamic equations are solved using the \textsc{athena++} radiation code. 

For the infrared radiation in the rapidly star-forming regions with wavelength $\lambda \geq 10\mu$m, the albedo of dust scattering is $\ll 1$ thus dust scattering is negligible compared to absorption (\citealt{Draine03}). We neglect dust scattering in the paper ($\kappa_s = 0$). For a Milky-Way-like dust-to-gas ratio, the infrared absorption opacity $\kappa_{aR}$ and $\kappa_{aP}$ on dust can be calculated by (see \citealt{KT12})
\begin{eqnarray}
(\kappa_{R},\kappa_{P})= (10^{-3/2},10^{-1}) \left( \frac{T}{10 \; \textrm{K}}\right)^2 \; {\rm cm^2 \; g^{-1}}.\label{opacity}
\end{eqnarray}
Equation (\ref{opacity}) is valid for the dust temperature $T\lesssim150\,$K, and $\kappa_{\rm R}$ becomes flat at the temperature range $150\,$K $\lesssim T \lesssim1000\,$K, then the dust reaches its sublimation temperature at $T\simeq 1000\,$K (\citealt{Semenov03}). In this paper we assume that the gas temperature is the same as the dust temperature in the clouds $T_{\rm dust}=T_{\rm gas}$. We denote $T_{c}$ as the cloud temperature and choose $T=T_{c}$ in equation (\ref{opacity}). The justification of this assumption is discussed in Section \ref{section_cloudmodels}.

\subsection{Cloud Models}\label{section_cloudmodels}

We define a constant infrared flux $F_*$ and a characteristic temperature
\begin{equation}
T_*=\left(\frac{F_*}{a_r c}\right)^{1/4},\label{dimensionless1}
\end{equation}
Following the argument in \cite{KT13}, the dust is considered to be initially in thermal equilibrium with the radiation field $T_{\rm dust}=T_*$. The cloud is considered to be in ``warm" starburst galaxies with a typical temperature $T_{\rm c}\sim 10 -100\,$K (\citealt{Gonzales04}; \citealt{Thompson05}; \citealt{Murray07}; \citealt{Andrews11}; \citealt{Scoville13}; \citealt{Vollmer17}). This is different from warm clouds in which photoheating balances radative cooling and have $T_{\rm c}\simeq 10^{4}\,$K. We assume that the cloud has an initial temperature of $T_{\rm c}=T_*=100\,$K for both dust and gas in the cloud, and this gives the fiducial radiation flux $F_* =5.6\times 10^{13}\,L_{\odot}\,$kpc$^{-2}$ with $T_{*,2}=T_*/100$\,K. This flux is comparable to some bright ultraluminous infrared galaxies (ULIRGs) with a star formation rate surface density $\Sigma_{\rm SFR}\sim 9.7 \times 10^{3}\,M_{\odot}$\,yr$^{-1}$\,kpc$^{-2}$ (\citealt{Kennicutt98}). In Section \ref{section_ULIRGs} we discuss other values of  $T_*$ and $F_*$ based on observations. As the cloud evolves, the radiation energy density $E_r$ in the cloud changes but the approximation $T_{c}\simeq T_r = (E_r/a_r c)^{1/4}$ holds. \cite{Davis14} performed alternative simulations with $\kappa_{R,P}\propto T_r ^{2}$ and compared to the standard simulations with $\kappa_{R,P}\propto T_c ^{2}$ but found similar results. 

Also, we define a characteristic acceleration
\begin{equation}
g_*=\frac{\kappa_{\rm R,*}F_*}{c},\label{dimensionless2}
\end{equation}
which measures the radiation force on the dust. In this work we consider two groups of simulations with different characteristic lengthscales. One is defined as ``large-scale" runs, in which the lengthscale is $h_* =0.1\,$pc. Another group is defined as ``small-scale" runs, which has a lengthscale of $h_*=c_{s,*}^{2}/g_*$, where $c_{s,*}$ is the characteristic sound speed. The small-scale $h_*$ is the pressure scale height similar to the scales in \cite{KT12,KT13} (see also \citealt{Davis14,Zhang17}), and small-scale runs are important to address the turbulence behavior of the clouds. We perform different scale runs to study whether different lengthscales of the clouds change the cloud evolution. Moreover, we take the cloud initial radius to be $r_*=50 h_*$, thus the large-scale cloud has a diameter of 10 pc, and the small-scale cloud has a diameter $\simeq 1\times 10^{-3}\,$pc for $T_* =100\,$K. Given different $h_*$, the characteristic timescale $t_*$ for large-scale runs is
\begin{equation}
t_{*}=\frac{h_*}{c_{s,*}} \approx 1.1\times10^{5}\,\textrm{yr}\label{time_large},
\end{equation}
while $t_*$ for small-scale runs is
\begin{equation}
t_{*}=\frac{h_*}{c_{s,*}} \approx 1.2\times10^{3}\,\textrm{yr}\label{time_small}.
\end{equation}
Note that physically $t_*=0.01 t_{\rm cc}= 0.01 r_*/ c_{s,*}$, where $t_{\rm cc}$ is the internal sound crossing time of the cloud. 

The initial infrared optical depth of a cloud along its diameter is $\tau_* = 2\kappa_{\rm R,*}\rho_* r_*$, where $\rho_*$ is the initial average density of the cloud. Observations show that the Na \textsc{I} carrying clouds have a typical column density $N_{\rm H}\sim 10^{21}\,$cm$^{-2}$ (e.g., \citealt{Martin05,Wofford13,Martin15}), which corresponds to $\tau_* \sim 0.01$. Another interesting consideration is the critical case $\tau_*=1$, in which the infrared flux just becomes opaque to the cloud. The case $\tau_*=1$ is still possible since it corresponds to a cloud column density $N_{\rm H}\sim2\times10^{23}\,$cm$^{-2}$ or $\sim 1.5\times10^{3}\,M_{\odot}\,$pc$^{-2}$, which is lower than the gas surface density in typical luminous infrared galaxies (LIRGs) and ULIRGs. Thus, in this paper we consider $\tau_*=0.01$ or 1 as the most interesting cases, but we also carry out simulations with $\tau_*=3$ and 10 for comparison. 

In our simulations the initial average density of the clouds is $\rho_*\sim 6.1\times 10^{3}\tau_*$ cm$^{-3}$ for large-scale runs, and $\rho_*\sim 5.4\times 10^{5}\tau_*$ cm$^{-3}$ for small-scale runs. \cite{Goldsmith01} studied that dust and gas can be thermally well coupled $T_{\rm dust}=T_{\rm gas}$ of $\rho_* \gtrsim 10^{4}$ cm$^{-3}$. Thus the assumption that dust and gas in the cloud has a same temperature approximately holds for $\tau_* \gtrsim 1$ cloud, but the thermally coupling breaks down for $\tau_*=0.01$. However, since the cloud is accelerated to be highly supersonic, the dynamics of the cloud is unlikely to be affected if we assume a different gas temperature from the dust temperature (\citealt{KT13}, Appendix A). 

Similar to \cite{Davis14}, an initial density perturbation with $\delta \rho /\rho$ randomly distributed between $-0.25$ to 0.25 is put into the clouds. The perturbation is somewhat arbitrary, in Section \ref{section_secondfactors1} we discuss the effect of cloud initial turbulent structure on cloud dynamics and evolution.

\begin{table*}[t]
\begin{center} 
Summary of Simulation Parameters

\begin{tabular} {lcllccllc}
\hline\hline
Run & dimensions & $\tau_*$ & $h_*$ & $\chi_0$ & $M_{\rm hot}$ & $[L_x \times L_z]/h_*$ & $N_x \times N_z$ & $\Delta z/h_*$\\
\hline
T0.01L       &  2D & 0.01    & 0.1 pc      & $10^{4}$ & -- & $400\times 1000$ & $400 \times 1000$ & 1\\
T1L       &  2D & 1    & 0.1 pc            & $10^{4}$ & -- & $400\times 1000$ & $400 \times 1000$ & 1 \\
T3L       &  2D & 3    & 0.1 pc            & $10^{4}$ & -- & $400\times 1000$ & $400 \times 1000$ & 1 \\
T10L       &  2D & 10    & 0.1 pc            & $10^{4}$ & -- & $400\times 1000$ & $400 \times 1000$ & 1 \\
T0.01S    &  2D & 0.01 & $c_{s,*}^{2}/g_*$ & $10^{4}$ & -- & $400\times 1000$ & $400 \times 1000$ & 1\\
T1S       &  2D & 1    & $c_{s,*}^{2}/g_*$ & $10^{4}$ & -- & $400\times 1000$ & $400 \times 1000$ & 1\\
T10S       &  2D & 10    & $c_{s,*}^{2}/g_*$ & $10^{4}$ & -- & $400\times 1000$ & $400 \times 1000$ & 1\\
\hline
T0.01L\_W &  2D & 1    & 0.1 pc            & $10^{2}$ & -- & $400\times 1000$ & $400 \times 1000$ & 1 \\
T1L\_W &  2D & 1    & 0.1 pc            & $10^{2}$ & -- & $400\times 1000$ & $400 \times 1000$ & 1 \\
\hline
T1LR1       &  2D & 1    & 0.1 pc            & $10^{4}$ & -- & $400\times 1000$ & $400 \times 1000$ & 1\\
T1LR2.5     &  2D & 1    & 0.1 pc            & $10^{4}$ & -- & $400\times 1000$ & $400 \times 1000$ & 1 \\
T1LR3       &  2D & 1    & 0.1 pc            & $10^{4}$ & -- & $400\times 1000$ & $400 \times 1000$ & 1 \\
\hline
T1L\_HR1  &  2D & 1    & 0.1 pc            & $10^{4}$ & -- & $400\times 1000$ & $800 \times 2000$ & 0.5 \\
T1L\_HR2  &  2D & 1    & 0.1 pc            & $10^{4}$ & -- & $400\times 1000$ & $800 \times 4000$ & 0.25$^{\rm a}$ \\
T1L\_LR1  &  2D & 1    & 0.1 pc            & $10^{4}$ & -- & $400\times 1000$ & $200 \times 500$ & 2 \\
T1L\_LR2  &  2D & 1    & 0.1 pc            & $10^{4}$ & -- & $400\times 1000$ & $100 \times 250$ & 4 \\
T0.01L\_HR   &  2D & 0.01    & 0.1 pc      & $10^{4}$ & -- & $400\times 1000$ & $200 \times 500$ & 2\\
T0.01L\_LR1   &  2D & 0.01    & 0.1 pc      & $10^{4}$ & -- & $400\times 1000$ & $200 \times 500$ & 2\\
\hline
T1L\_3D1  &  3D & 1    & 0.1 pc            & $10^{4}$ & -- & $400^{2} \times 1000$ & $200^{2} \times 500$ & 2 \\
T1L\_3D2  &  3D & 1    & 0.1 pc            & $10^{4}$ & -- & $400^{2} \times 1000$ & $100^{2} \times 250$ & 4 \\
\hline
H1 &  2D & --    & 2.8e-4 pc            & $10^{4}$ & $5$ & $400\times 1000$ & $400 \times 1000$ & 1 \\
H2 &  2D & --    & 1.1e-3 pc            & $10^{4}$ & $10$ & $400\times 1000$ & $400 \times 1000$ & 1 \\
H3 &  2D & --    & 4.5e-3 pc$^{\rm b}$            & $10^{4}$ & $20$ & $400\times 1000$ & $400 \times 1000$ & 1 \\
\hline \hline
\end{tabular}
\end{center}
\caption{Notes: Parameter Definitions: $\tau_*$ is the initial infrared optical depth of the clouds, $h_*$ is the characteristic lengthscale, $\chi_0$ is the initial density ratio of the clouds to the background medium, $M_{\rm hot}$ is the Mach number in hot flows given by $V_{\rm hot}/c_{s,\rm hot}$, $L_x$ and $L_z$ are the length of the computational box along $x$ and $z$ directions in unit of $h_*$, and $N_x$ and $N_z$ are the grid zones along $x$ and $z$ directions. $^{\rm a}$For T1L\_HR2 run $\Delta z/h_* = 0.25$ but $\Delta x/h_*=0.5$. $^{\rm b}$For H1, H2 and H3 runs the lenthscale is given by $\tau_* h_*/(2\kappa_{R,*}\rho_*)$ where $\rho_*=\chi_0 \rho_{\rm hot}$, $h_*=0.1\,$pc, $\chi_0=10^4$ and $\rho_{\rm hot}$ is given by equation (\ref{windmomentun}).}\label{tab_parameter}
\end{table*}

\subsection{Background Medium}\label{section_background}

If the cloud is not confined by the background pressure, it expands approximately at its sound speed. \cite{Thompson15} showed that the semi-analytic model for cloud expansion and acceleration in a vacuum radiation field. However, it is more likely that the cloud is initially in approximate thermal pressure equilibrium with its surrounding medium (e.g., \citealt{Spitzer68}; \citealt{SH07}; \citealt{SB15}; \citealt{SR17}). In general, a hot ionized ISM has a typical temperature $\sim  10^{6}$\,K, and a warm ionized ISM has $\sim  10^{4}$\,K. We assume that the cloud is embedded in the hot or warm ISM and in thermal pressure equilibrium with the ISM. We introduce the initial density ratio of the cloud to the background medium $\chi_0  = \rho_*/\rho_{\rm bkgd} = T_{\rm bkgd}/T_*$, and consider two cases, $\chi_0  = 10^{4}$ and $\chi_0  = 10^{2}$, which corresponds to $T_{\rm bkgd}=10^{6}\,$K and $10^{4}\,$K respectively. Here, we have dropped the dependence on the mean molecular weight and take $\mu=1$. In Section \ref{section_secondfactors1} we consider the effect of molecular weights of the cloud and background. Also, we only include dust heating and cooling via radiation in our simulations, thus the background is adiabatic. We discuss the relevance of ionized background heating and cooling and the possibility that the cloud is not in pressure equilibrium with the background in Sections \ref{section_cooling} and \ref{section_secondfactors2}. 

Thermal conduction can also be important due to temperature gradient between the cloud and background. There may be a large thermal conductive flux from the background to the cloud. In the case of saturated evaporation, \cite{Cowie77} showed that the cloud would be evaporated on a timescale (see also \citealt{Faucher12}) 
\begin{equation}
t_{\rm evap}\sim 1.5\times 10^{9}\,\textrm{yr}\,N_{\rm H,23}T_{\rm c,2}^{-5/6}n_{\rm c,3}^{-1},
\end{equation}
where $N_{\rm H,23}=N_{\rm H}/10^{23}\,$cm$^{-2}$, $T_{\rm c,2}=T_{\rm c}/100\,$K and $n_{\rm c,3}=n_{\rm c}/10^{3}\,$g cm$^{-3}$. Thus, we have
\begin{eqnarray}
t_{\rm evap}&\sim& 5.1\times 10^{8}\tau_*^{1/6}\,\textrm{yr}\qquad\textrm{for large-scale runs}\\
&\sim& 5.7\times 10^{6}\tau_*^{1/6}\,\textrm{yr}\qquad\textrm{for small-scale runs},
\end{eqnarray}
both of which are much longer than $t_*$ ($t_{\rm evap}\gg t_*$). As a result, thermal conduction is not important based on the simply analytic estimate. Recent numerical simulations of clouds in hot galactic winds showed that thermal conduction may cause cold clouds to evaporate (e.g., \citealt{BS16}; \citealt{Waters16}), but it has also been argued that conduction may be suppressed either by magnetic fields or plasma instabilities (\citealt{McCourt16}). Nevertheless, in this work we neglect the effect of thermal conduction.

\subsection{Reduced Speed of Light Approximation}\label{section_reduced_speed1}

If we use $h_*$ as the unit for length, $c_{s,*}$ as the unit for velocity, $t_* = h_*/c_{s,*}$ as the unit for time, $a_r T_*^{4}$ as the unit for $E_r$ and ${\sf P}_r$, and $c a_r T_*^{4}$ as the unit for $\mathbf{F}_r$, the RT equation (\ref{RTequation}) can be dimensionalized as
\begin{equation}
\frac{\partial \hat{I}}{\partial \hat{t}}+\Crat \bn\cdot\bfnabla \hat{I}= \Crat S(\hat{I},\bn) \label{radsource1},
\end{equation}
where $\Crat = c / c_{s,*}$ is the ratio between the speed of light and the characteristic sound speed, and $\hat{I}$ and $\hat{t}$ are the dimensionless intensity and time. Since the fluid velocities are small compared to the speed of light to the order of $\mathcal{O}\left(v/c\right)$, one can reduce the speed of light and introduce a dimensionless parameter $\tilde\Crat$, where $\tilde\Crat \ll \Crat$ but $\tilde\Crat \gg 1$. The dimensionless RT equation (\ref{radsource1}) can be modified to
\begin{equation}
\frac{\partial \hat{I}}{\partial \hat{t}}+\tilde\Crat \bn\cdot\bfnabla \hat{I}= \tilde\Crat S(\hat{I},\bn) \label{reduced_radsource1}.
\end{equation}
We define the reduction factor $\Rrat = \tilde\Crat/ \Crat $ and take $\Rrat =10^{-2}$ as the fiducial value throughout the paper. We also test other reduction factors $\Rrat=10^{-1}$, $10^{-2.5}$ or $10^{-3}$ in Section \ref{section_reduced_speed2}. 

On the other hand, the dimensinonless hydrodynamic momentum and energy equations can be written as  
\begin{eqnarray}
\pdif{\left(\hat{\rho}\mathbf{\hat{v}}\right)}{\hat{t}} + \mathbf{\nabla} \cdot \left( \hat{\rho}\mathbf{\hat{v}} 
\mathbf{\hat{v}} + {\sf \hat{P}}\right) &=& \hat{\rho} \mathbf{\hat{g}} - \Prat\mathbf{S}_r(\mathbf{P}), \\
\pdif{\hat{E}}{\hat{t}} + \mathbf{\nabla} \cdot \left(\hat{E} \mathbf{\hat{v}} + {\sf \hat{P}} \cdot \mathbf{\hat{v}}\right) &=& 
\hat{\rho} \mathbf{\hat{g}} \cdot \mathbf{\hat{v}} -\Prat \Crat S_r(E)\label{reduced_hydro3}.
\end{eqnarray}
Comparing to equations (\ref{hydroequation2}) and (\ref{hydroequation3}), the quantities denoted with hat ( $\hat{}$ ) are dimensionlized, and the parameter $\Prat$ is defined as $\Prat=a_r T_*^{4}/(\rho_* c_{s,*}^{2})$ (\citealt{Jiang12}). We solve dimensionless equations instead of equations (\ref{hydroequation1}), (\ref{hydroequation2}), (\ref{hydroequation3}) and (\ref{RTequation}), and hereafter skip the hat denotations for all the quantities. Note that $\Crat$ in hydrodynamic equation (\ref{reduced_hydro3}) is not reduced. Importantly, the reduced speed of light only appears in the RT equation (see \cite{Gnedin16} for a detailed discussion). More details of the reduced speed of light approximation is discussed in Section \ref{section_reduced_speed2} and Appendix \ref{append_reduced_speed}.

\subsection{Initial Setup}\label{section_initial}

We perform a series of 2D and 3D simulations on a Cartesian grid. In 2D runs $x$-coordinate is the horizontal direction, and $z$-coordinate is the vertical direction. In 3D runs we add $y$-coordinate as another horizontal direction.  Flux $F_*$ is injected at the bottom of the vertical direction as the radiation boundary. Other radiation boundary and all the hydrodynamic boundaries are set up as outflows. In all simulations, the computational box covers ($-200\,h_*$, $200\,h_*$) in $x$ direction and ($-500\,h_*$, $500\,h_*$) in $z$ direction, for 3D runs the box also covers ($-200\,h_*$, $200\,h_*$) in $y$ direction. The cloud is located at the center of the computational domain $(x,z)=(0,0)$.


The mass-weighted cloud mean velocity during the cloud acceleration is given by
\begin{equation}
V_{\rm mean}=\langle v_z \rangle =\frac{1}{M_{c}}\int_{V_{c}}\rho v_{z} dV.
\end{equation}
The ``flying distance" of the irradiated cloud along the vertical direction can be defined at the mass center of the cloud
\begin{equation}
z_{\rm c}=\int_{0}^{t}V_{\rm mean} dt\label{distance}.
\end{equation}
Since $z_{\rm c}$ is much larger than the size of the box in most simulations, we adopt the cloud-following scheme to keep the cloud always at the center of the box. In cloud-following frame the cloud mean velocity along the vertical direction is always zero, and the background medium has a velocity of $-V_{\rm mean}$. The vertical coordinate $z$ is replaced by $z+z_{c}$ instead, but we still denote it as $z$.  

Table \ref{tab_parameter} summarizes simulation parameters for our 2D and 3D radiation hydrodynamic runs with large-scale and small-scale of $h_*$ and a range of cloud initial optical depth $\tau_*$, reduction factor $\Rrat$, density ratio $\chi_{0}$ and resolutions. T0.01L, T1L, T3L and T10L denotes large-scale (L) runs with $\tau_*=0.01$, 1, 3 and 10 respectively. T0.01S, T1S, T10S are small-scale (S) runs with $\tau_*=0.01$, 1 and 10. T0.01L\_W and T1L\_W are runs with $\chi_{0}=10^{2}$ so they fall to the case of warm (W) background with $T_{\rm bkgd}=10^{4}\,$K. Since we choose a fiducial reduction factor $\Rrat=10^{-2}$ for all above runs, we investigate the reduced speed of light approximation with runs T1LR1, T1LR2.5 and T1LR3, which correspond to $\Rrat=10^{-1},10^{-2.5}$ and $10^{-3}$ respectively. Moreover, T1L\_HR1, T1L\_HR2, T1L\_LR1 and T1L\_LR2 are runs to test the effects of spatial resolution compared to T1L, and T0.01L\_HR and T0.01L\_LR1 and T0.01L\_LR2 are also carried out to compared to T0.01L. T1L\_3D1 and T1L\_3D2 are 3D runs with $\tau_*=1$ and different resolutions. Next we introduce runs H1, H2 and H3.

\subsection{Hot Wind vs. Radiation}\label{section_hotwind1}

Hot galactic winds may be driven by supernova explosions in star-forming and starburst galaxies (\citealt{CC85}). Besides radiation pressure on dust, it has been proposed that cold clouds may also be accelerated by ram pressure of the hot flow to observed velocities (\citealt{SS00}; \citealt{Murray07};\citealt{Cooper08, Cooper09}; \citealt{Fujita09}). However, recently both analytic calculations and numerical simulations shows that clouds are more likely to be quickly destroyed by shearing and Kelvin-Helmholtz instabilities before its fully acceleration to the velocity of the hot flow (\citealt{SB15}, \citealt{BS16}, \citealt{SR17}; \citealt{Zhang17}). 

In this paper, we also compare the dynamics of cloud acceleration in the environment of hot wind and in an infrared radiation field. \cite{HT17} estimate that the momentum injection rate of a hot wind is comparable to the expectation from radiation pressure in star-forming galaxies. We run simulations with a hot wind boundary condition to replace the radiation boundary. In order to compare to a fiducial radiation run, we set the momentum injection of hot wind $\rho_{\rm hot}V_{\rm hot}^{2}$ to be equal to the flux momentum $F_*/c$, where $\rho_{\rm hot}$ and $V_{\rm hot}$ are the density and velocity of the hot wind respectively. For simplicity, we do not adopt the widely used analytic models of hot wind given by \cite{CC85}, but take Mach number of the hot wind $M_{\rm hot}=V_{\rm hot}/c_{s,\rm hot}$ as a parameter. We set $M_{\rm hot}=5$, 10 and 20, and the density of hot wind $\rho_{\rm hot}$ is given by
\begin{equation}
\rho_{\rm hot}=\frac{F_*}{c V_{\rm hot}^{2}}=\frac{a_r T_*^{4}}{\chi_0 c_{s,*}^{2}M_{\rm hot}^{2}}.\label{windmomentun}
\end{equation}
H1, H2, H3 in Table \ref{tab_parameter} show the initial conditions for the simulations of cloud in a hot flow. The cloud has an initial temperature $T_*=100\,$K, and density perturbation $\delta\rho/\rho$ is the same as in radiation runs, and the cloud is initially in thermal equilibrium with the hot medium with $\chi_{0}=10^{4}$. The lengthscale $h_*$ for the cloud-hot flow runs are adjusted so that the initial column density of the cloud is exactly the same as that in radiation runs with $\tau_*=1$. We find that $h_*=2.8\times 10^{-4}$ pc, $1.1\times 10^{-3}$ pc and $4.5\times 10^{-3}$ pc for H1, H2 and H3 respectively. 

Some other runs are discussed in Section \ref{section_discussion}. 


\begin{figure*}[t]
\centerline{\includegraphics[width=17.5cm]{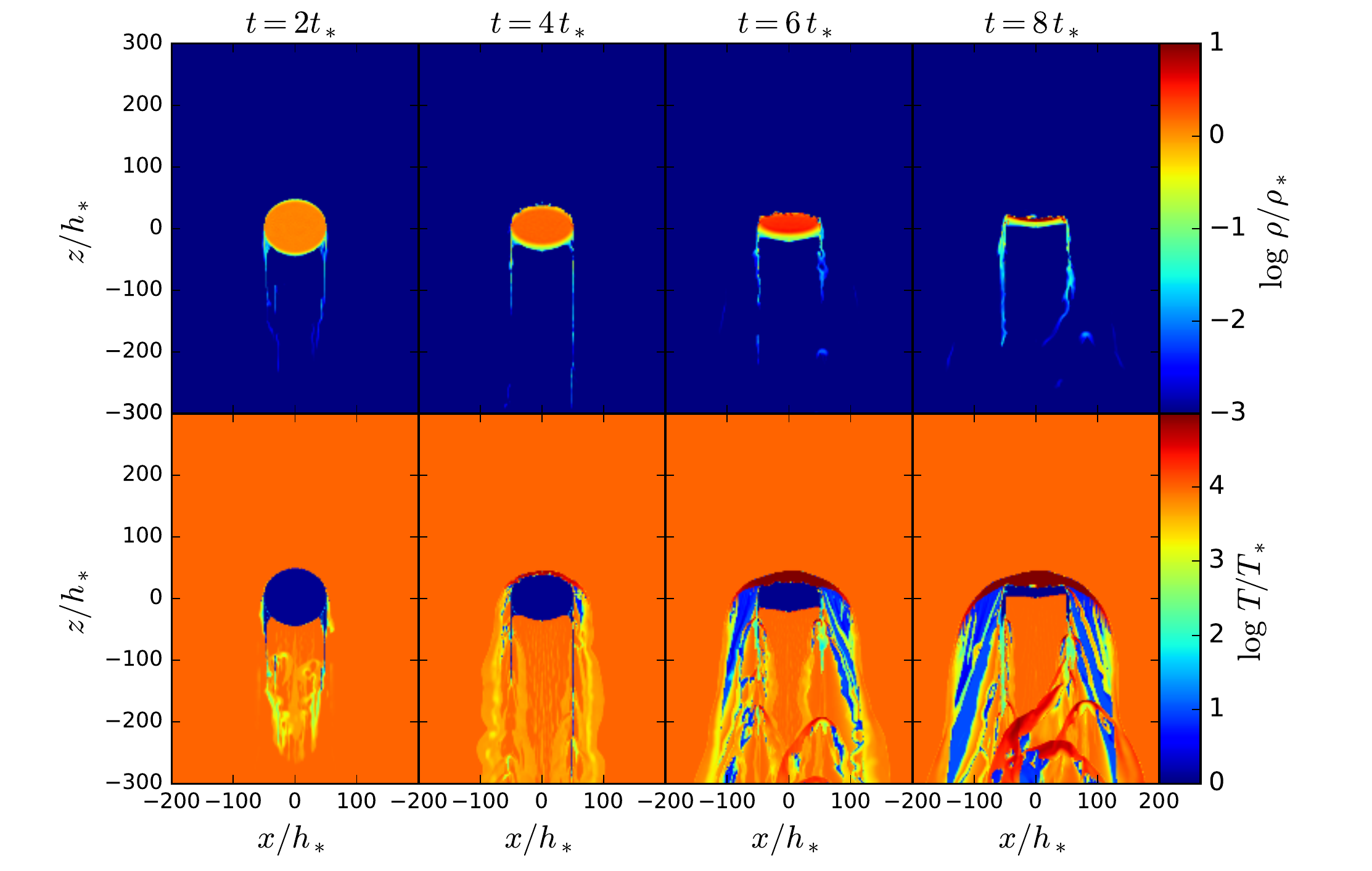}}
\caption{Density and temperature snapshots in run T0.01L, which is the large-scale (L) run with initial optical depth $\tau_* =1$. The lengthscale in large-scale runs is $h_* = 0.1\,$pc, and time unit $t_* \approx 1.1\times 10^{5}\,$yr (see equation \ref{time_large}).}\label{fig_T001L}
\end{figure*}
\begin{figure}[t]
\centerline{\includegraphics[width=9cm]{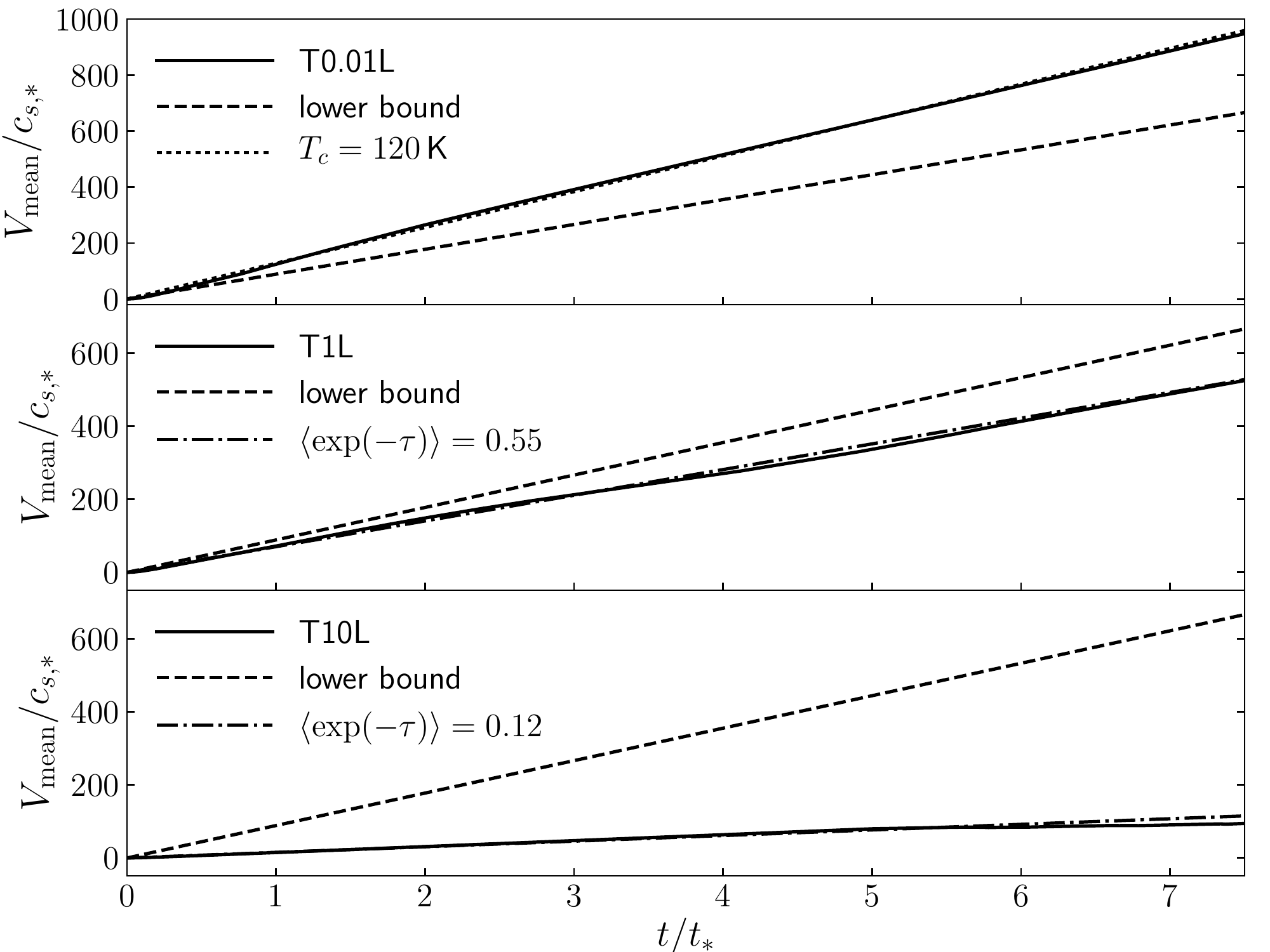}}
\caption{Cloud mean velocity $V_{\rm mean}$ for runs T0.01L, T1L and T10L, which correspond to large-scale runs with $\tau_*=0.01$, 1 and 10 respectively. The lower bound (dashed lines) for {\it optical thin cloud} acceleration is given by equation (\ref{analytic}). The fitting line in upper panel is also given by equation (\ref{analytic}) with $T_{c}=120\,$K, and the fitting lines in middle and lower panels are given by equation  (\ref{analytic2}) with the averaged attenuation factor $\langle {\rm e}^{-\tau} \rangle=0.55$ and 0.12 respectively.}\label{fig_vmean_a}
\end{figure}
\begin{figure*}[t]
\centerline{\includegraphics[width=18cm]{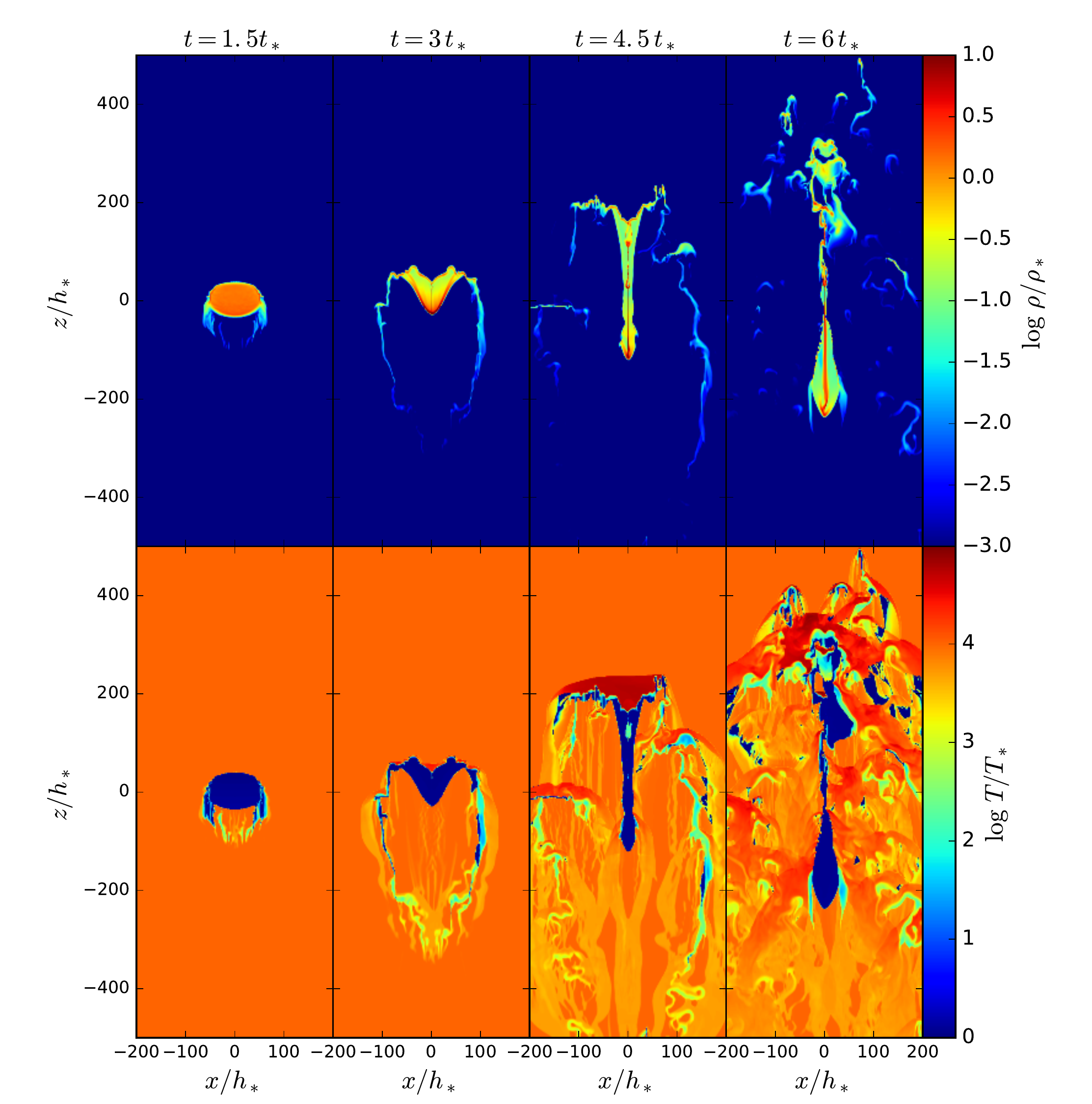}}
\caption{Density and temperature snapshots in run T1L, which is the large-scale run with $\tau_*=1$. The lengthscale $h_*$ and time unit $t_*$ are the same as in Figure \ref{fig_T001L}.}\label{fig_T1L}
\end{figure*}
\begin{figure}[t]
\centerline{\includegraphics[width=8.5cm]{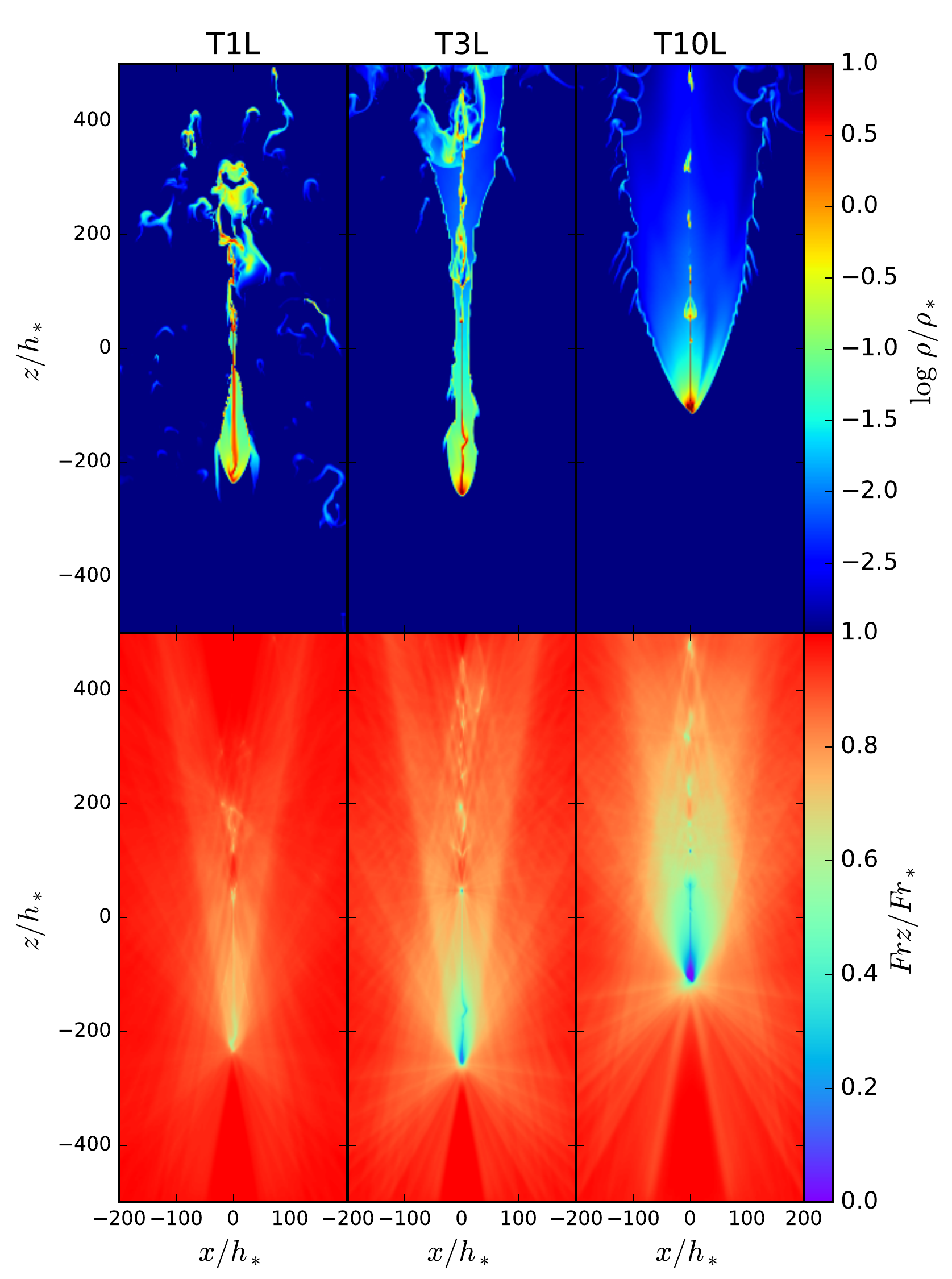}}
\caption{Density and $z-$component flux $F_{rz}$ snapshots from runs T1L, T3L and T10L at $t=6\,t_*$.}\label{fig_thick}
\end{figure}
\begin{figure}[t]
\centerline{\includegraphics[width=8.5cm]{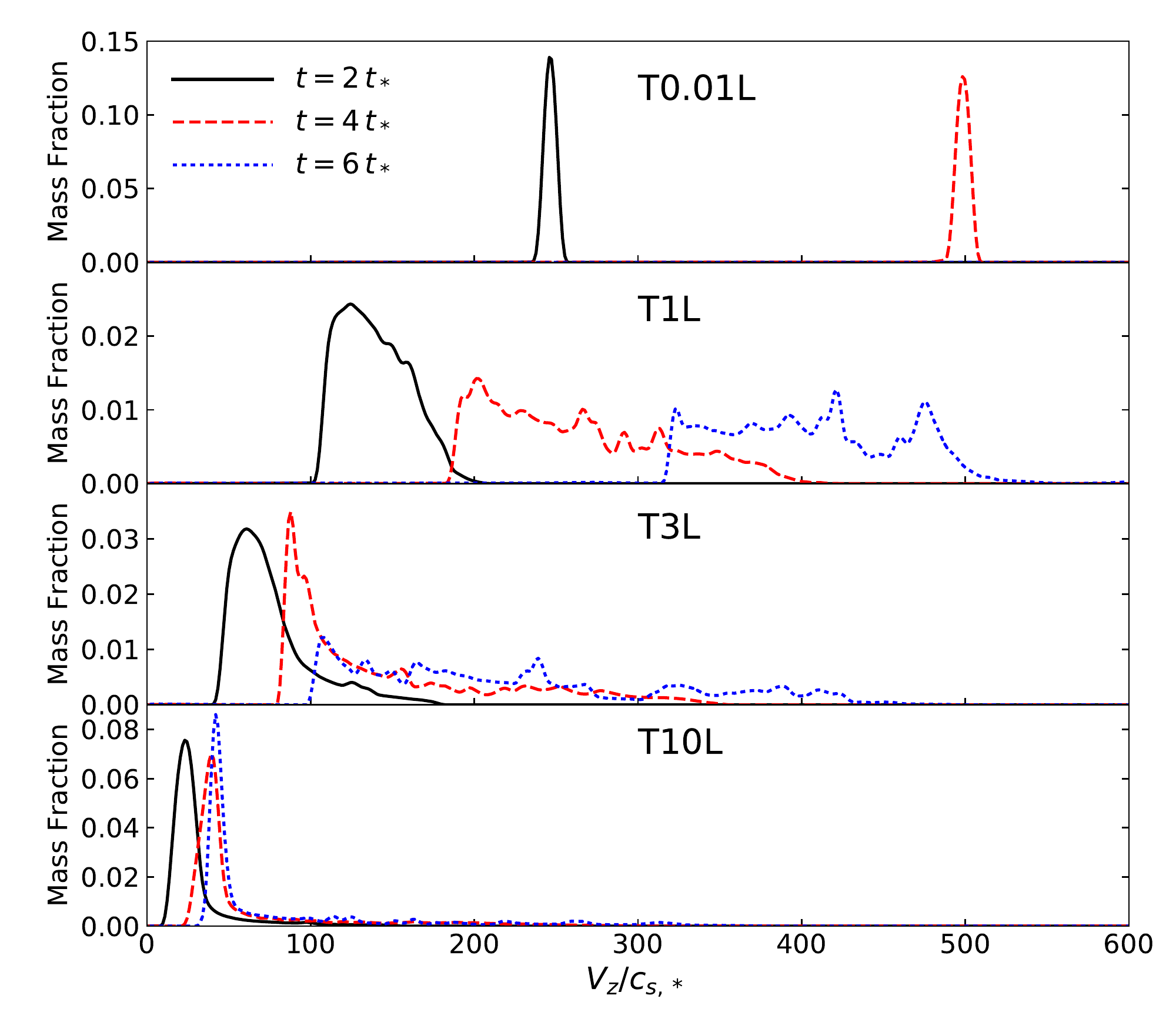}}
\caption{Velocity probability distribution function (PDF) for runs T0.01L, T1L, T3L and T10L at $t=2\,t_*$, $4\,t_*$ and $6\,t_*$.}\label{fig_dist}
\end{figure}

\section{Simulation Results}\label{section_results}

\subsection{Large-Scale Runs: Optically Thin Case}\label{section_thin}

We first run T0.01L, which corresponds to the large-scale run with $\tau_*=0.01$. For simplicity, we assume the flux is constant along the vertical direction, and turn off the gravity. Figures \ref{fig_T001L} shows density and temperature snapshots from T0.01L. 
The cloud develops a significant ``pancake" structure at $t\gtrsim 4\,t_*$ elongated along the horizontal direction. The vertically upward radiation pressure on dust which accelerates the cloud and the downward gas pressure from the background and the ram pressure due to the differential velocity between the front of the cloud and background combine together to squeeze the cloud along the vertical direction.
Also, we find a downward ``tail" structure forms at both sides of the cloud, which is caused by ram pressure of the hot background shearing the edge of the cloud. The lower panel of Figure \ref{fig_T001L} shows that the core of the cloud remains cold and dense during its acceleration. The interface between the front of the cloud and the background is heated to as high as  $T/T_*\sim 10^{5}$ (i.e., $T\sim 10^{7}\,$K).  The low-density gas in the tails of the cloud behind the pancake structure is mixed up with gas temperature to $\sim 500-1000\,$K, in which dust is sublimated. However, most mass of the cloud is still distributed in the pancake structure which is heated by radiation flux to a slightly higher temperature $T_{c}\sim 120\,$K with $\rho \sim 1-50\,\rho_*$. The irradiated cloud continues to be accelerated by radiation pressure on dust.  

The upper panel of Figure \ref{fig_vmean_a} shows $V_{\rm mean}$ of the cloud in T0.01L. The cloud is almost linearly accelerated with time, suggesting that shredding and turbulence in the cloud has little impact on cloud acceleration.  For an optically thin cloud, radiation flux penetrates the entire cloud without attenuation, the cloud can be accelerated uniformly as a whole, thus the cloud mean velocity can be simply estimated by
\begin{eqnarray}
V_{\rm mean} &=& \int \frac{\kappa F}{c}dt \nonumber\\
&\sim & V_*  \left(\frac{\textrm{min}\{T_c,150\,\textrm{K}\}}{T_*}\right)^{2}\left(\frac{t}{t_*}\right),\label{analytic}
\end{eqnarray}
Note that the dust opacity is $\propto T_{c}^{2}$ at $T_{c}<150\,$K (equation \ref{opacity}), and becomes fairly flat at $150\,$K$<T_{c}<1000\,$K. Equation (\ref{analytic}) gives a linearly increased $V_{\rm mean}$ with time $t$, where $V_* = \kappa_{\rm R} F_* t_*/c = g_* t_*$ is the characteristic cloud velocity determined by the initial cloud temperature, radiation flux and the cloud characteristic lengthscale. Combining equations (\ref{dimensionless1}) and (\ref{dimensionless2}), we have $V_* \simeq 89 c_{s,*}$ for large-scale runs. Given $T_{c}\sim 120 \,$K in the cloud, we obtain $V_{\rm mean}\sim1.4\,V_* (t/t_*)$ from equation (\ref{analytic}), which is well consistent with the upper panel of Figure \ref{fig_T001L}. 

We stop the run at $t=8t_*$ when the cloud reaches $V_{\rm mean}\sim 10^{3}c_{s,*}$ ($\sim 920\,$km s$^{-1}$). However, note that the assumption of a constant flux may break down if the flying distance of the cloud $z_{c}$ is comparable to the size of the galaxy. The cloud flying distance  $z_{c}$ can be estimated by equation (\ref{distance}) that $z_{\rm c}=0.7 V_* t_* (t/t_*)^{2}\simeq 6.2\,$pc $(t/t_*)^{2}$. We have $z_{c}>R$ if 
\begin{equation}
t>5.7\,t_{*}R_{\rm 200pc}^{1/2},
\end{equation} 
with $R=200\,$pc$R_{\rm 200pc}$ being the size of the galaxy, thus the geometry of the galaxy becomes important to decelerated the cloud. In Section \ref{section_secondfactors2} we take into account the effect of gravity and varying flux along the height.

\begin{figure}[t]
\centerline{\includegraphics[width=8.5cm]{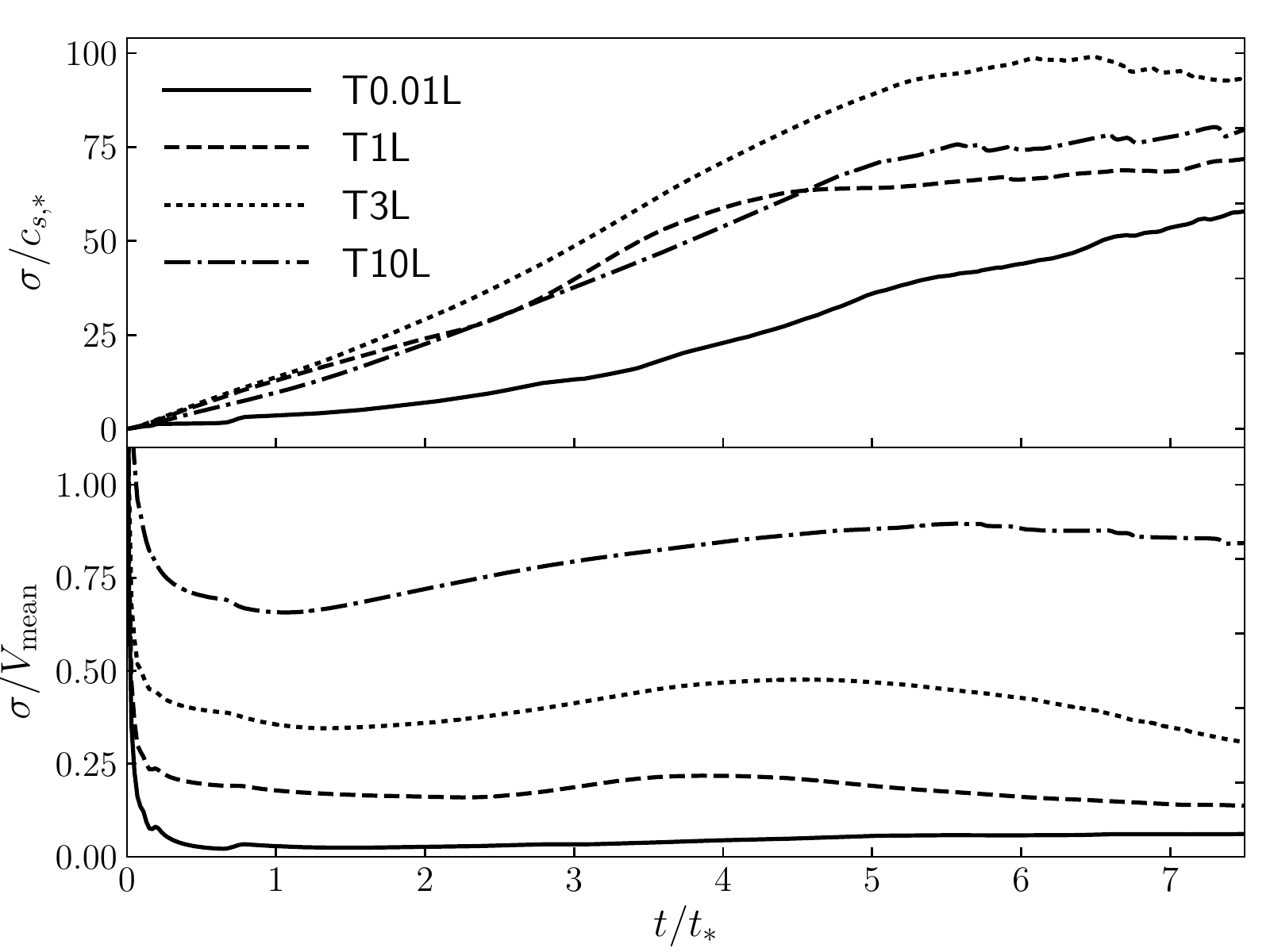}}
\caption{Velocity dispersion in the clouds $\sigma/c_{s,*}$ (upper panel) and the ratio $\sigma/V_{\rm mean}$ for runs T0.01L, T1L, T3L and T10L. }\label{fig_sigmaL}
\end{figure}
\begin{figure*}[t]
\centerline{\includegraphics[width=18cm]{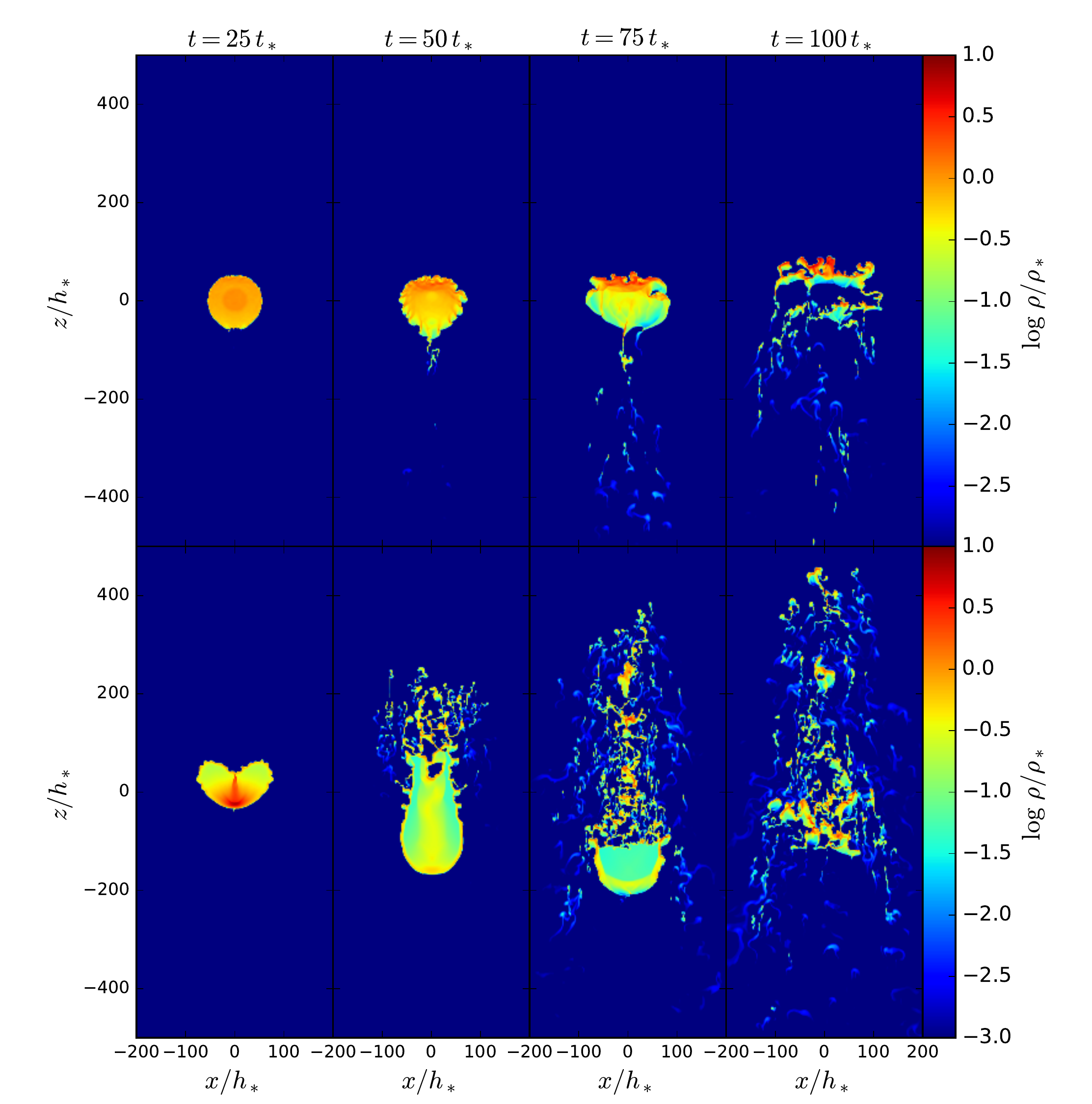}}
\caption{Density snapshots in runs T0.01S (upper panels) and T1S (lower panels), which are the small-scale (S) runs with $\tau_* =0.01$ and 1 respectively. The lenthscale for small-scale runs is $h_*=c_{s,*}^{2}/g_*$, and the time unit $t_*\approx 1.2\times 10^{3}\,$yr (see equations \ref{dimensionless2} and \ref{time_small}).}\label{fig_SR}
\end{figure*}
\begin{figure}[t]
\centerline{\includegraphics[width=8.5cm]{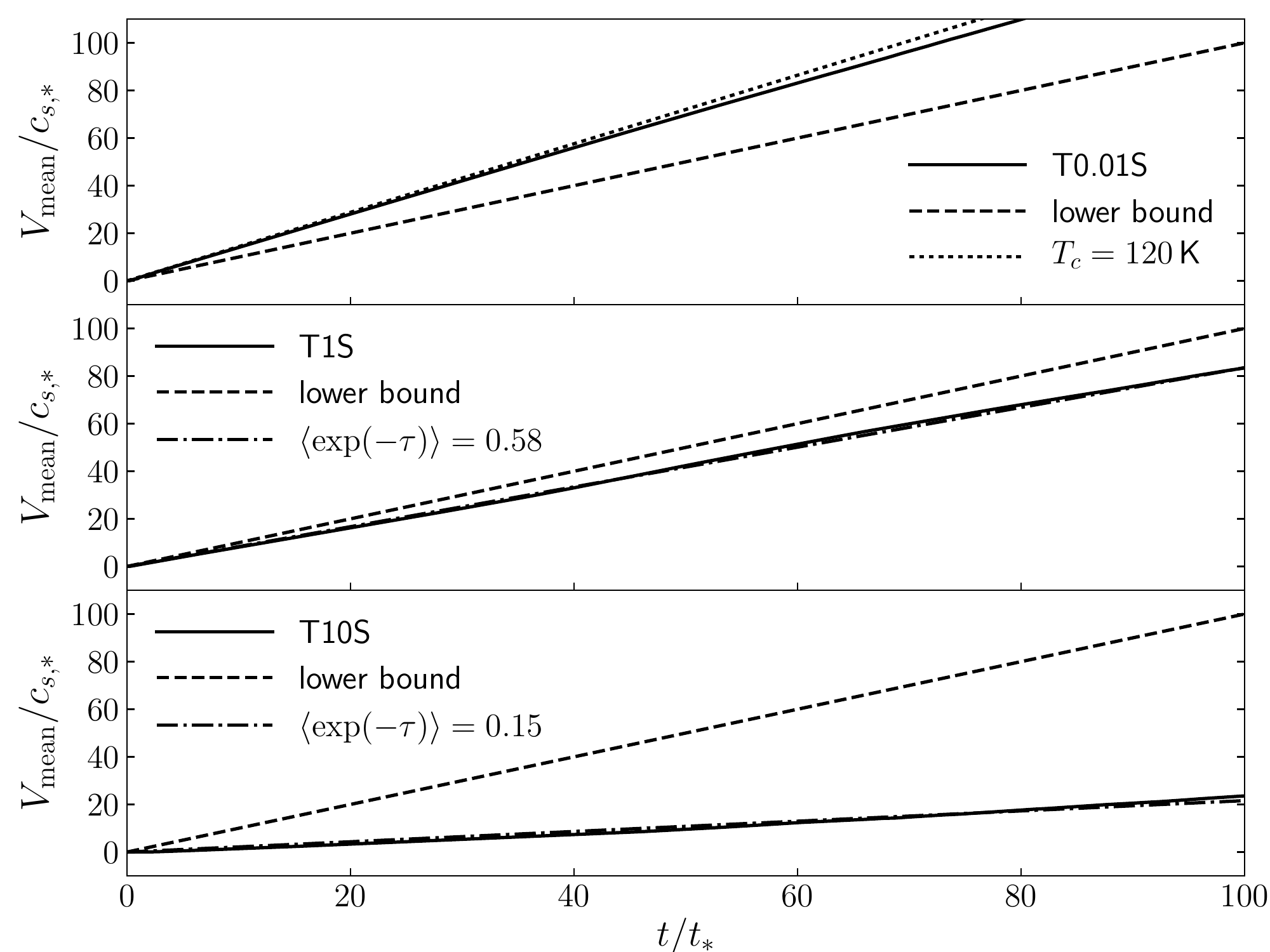}}
\caption{Cloud mean velocity $V_{\rm mean}$ for runs T0.01S (upper panel), T1S (middle panel) and T10S (lower panel), which correspond to the small-scale runs with $\tau_* =0.01$, 1 and 10 respectively. The lower bound (dashed lines) for {\it optical thin cloud} acceleration is given by equation (\ref{analytic}), also the lines denoted by $\langle \exp(-\tau) \rangle$ are plotted based on equation (\ref{analytic2}).}\label{fig_vmean_b}
\end{figure}

\subsection{Large-Scale Runs: Optically Thick Case}\label{section_thick}

We first perform the T1L run, which corresponds to the large-scale run with $\tau_*=1$, and take T1L as the fiducial run. Figure \ref{fig_T1L} shows density and temperature snapshots from T1L. The cloud is initially similar to the case of T0.01L with a pancake structure elongated horizontally and long tails behind the cloud at $t=1.5\,t_*$. However, the evolution of the irradiated cloud becomes quite different from T0.01L at late time. The cloud has optical depth $\tau_*=1$ along the vertical diameter, but the optical depth is lower at the outside region of the cloud. Since the radiation flux is attenuated by a factor of ${\rm e}^{-\tau}$ and distributed inhomogeneously inside the cloud, the cloud can no longer be accelerated uniformly as a whole. Different radiation force due to different flux inside the cloud produced a shear in the cloud -- the outside region of the cloud at both sides is accelerated faster than the center of the cloud. Most mass with lower velocity is dragged behind and accumulated at the bottom of the cloud. The differential velocity between the cloud outside and central regions eventually stretches out the cloud, leading to a filamentary structure elongated along the vertical direction. The gas in the filamentary structure is much denser than the outside region of the cloud. We find that the peak value of density in the central region increases to $\rho \sim 10-20\rho_*$ at $t=4.5\,t_*$ and slightly decreases to $\rho \sim 1-10\rho_*$ at $t=6\,t_*$. 

We also carry out other two large-scale runs T3L ($\tau_*=3$) and T10L ($\tau_*=10$). The mean velocity for optically thick cloud can be estimated by
\begin{equation}
V_{\rm mean}\sim V_* \langle {\rm e}^{-\tau} \rangle \left(\frac{\textrm{min}\{T_c,150\,\textrm{K}\}}{T_*}\right)^{2}\left(\frac{t}{t_*}\right),\label{analytic2}
\end{equation}
where $\langle {\rm e}^{-\tau} \rangle=\int \rho {\rm e}^{-\tau}dt dV /M_{c}t$ is the mass and time averaged attenuation factor, which measures the impact of optical depth on cloud acceleration. The middle and lower panels in Figure \ref{fig_vmean_a} shows $V_{\rm mean}$ from two runs T1L and T10L. We find that $T_{c}\sim120\,$K still holds for optically thick clouds. The values of $V_{\rm mean}$ can be approximately given by $\langle {\rm e}^{-\tau} \rangle\simeq0.55$ in equation (\ref{analytic2}) for T1L and $\langle {\rm e}^{-\tau} \rangle\simeq0.12$ for T10. Also, from T3L we find that $\langle {\rm e}^{-\tau} \rangle\simeq0.31$ for $\tau_*=3$, which is the intermediate case between T1L and T10L. Combining results from T1L to T10L, we give an empirical formula to fit the attenuation factor that 
\begin{equation}
\langle {\rm e}^{-\tau} \rangle = \exp(0.92-1.5\tau_*^{0.3})\label{taufitting1},
\end{equation}
which holds for $\tau_* \geqslant  1$. If we combine $\epsilon= \langle {\rm e}^{-\tau} \rangle (T_c/T_*)^2$ as a new parameter and $V_c = \epsilon V_* (t/t_*)$ for both optically thin and optically thick clouds, we obtain
\begin{equation}
\epsilon \simeq \textrm{min}\{1,\exp(1.3-1.5\tau_*^{0.3})\}\label{taufitting2},
\end{equation}
higher $\tau_*$ gives a slower acceleration. 

Figure \ref{fig_thick} shows snapshots of density and $z-$component radiation flux $F_{rz}$ for T1L, T3L and T10L at a same time $t=6\,t_*$. T3L also shows a clear filamentary structure stretched vertically, which is similar as in T1L. Cloud in T10L has the largest initial optical depth, radiation only blows away the outside region of the cloud at $t=6\,t_*$, but the core of the cloud is moving slowly. We find that a filamentary structure is eventually formed at $t\sim 8 t_*$ in T10L. Furthermore, the flux $F_{rz}$ is mainly attenuated in the densest region of the cloud and decreases to $F_{rz}\sim 0.1 F_{r,*}$ in the bottom of the cloud in T3L, and $F_{rz}\sim 0$ in T10L. On the contrary, we find that $F_{rz}$ in the run for optically thin cloud T0.01L is uniform with $F_{rz} \simeq F_{r,*}$ everywhere. Higher optical depth means heavier flux attenuation inside the cloud.  Note that this result is in contrast to the dusty shell acceleration (e.g., \citealt{KT12, KT13, Davis14, ZD17}), in which higher shell optical depth leads to a higher acceleration. The key difference between them is that the radiation flux can be well trapped inside a shell and the momentum coupling between the radiation field and the shell correlates with the shell optical depth, but the flux can easily escape from a cloud. As shown in Figure \ref{fig_thick}, the re-radiated flux from the cloud escapes away from the cloud, although the re-radiated flux profiles show a phenomena referred to as ``ray-effects", which correspond to unphysical anisotropies in the radiation field due to the angular resolution. Ray-effects have been discussed in the literature (e.g., \citealt{Larsen08}; \citealt{Finlator09}). The effects of angular resolution were explored in the appendix of \cite{Davis14} in which the convergence was found even when ray-effects were presented.

It is also important to study the spread/dispersion of velocities about the mean velocities for the clouds. Figure \ref{fig_dist} compares mass-weighted velocity probability distribution functions (PDFs) in the vertical direction for four runs from T0.01L to T10L. The PDFs for all runs shift to higher velocity of $v_{z}$, but the cloud with lowest optical depth $\tau_*\ll 1$ gains the highest acceleration but the tightest extension of velocities. This is consistent with the result that optically thin cloud is accelerated as a whole. The velocity distribution becomes more extended in T1L with time, and $v_{z}$ spreads from $V_z \sim 320\,c_{s,*}$ to $\sim 550\,c_{s,*}$ at $t=6\,t_*$. The velocity distribution becomes even more extended in T3L at $t=6\,t_*$, from $V_{z}\sim 100\,c_{s,*}$ to $\sim 450\,c_{s,*}$ at $t=6\,t_*$. The shape of the PDF for T10L is somewhat similar to T0.01L, however, T10L shows a slowest cloud acceleration, and the PDF of T10L shows a tailed profile extending to $V_z\sim 300 c_{s,*}$. A fraction of the cloud in T10L has been already accelerated to a high velocity, while the main body still remains a low velocity. 

Another quantity to measure the cloud velocity spread or turbulence is the mass weighted velocity dispersion of the cloud $\sigma$, which is given by
\begin{equation}
\sigma_{i}^2 = \frac{1}{M} \int \rho (v_i-\langle v_i\rangle)^2 dV,
\end{equation}
where $i$ has $x$ and $z$ components for 2D runs, and $x$, $y$, $z$ components for 3D runs, and the total velocity dispersion is $\sigma=\sqrt{\Sigma\sigma_i^{2}}$. Figure \ref{fig_sigmaL} shows the properties of velocity dispersion in four runs. We find that T3L shows the largest velocity dispersion that $\sigma\sim 100\, c_{s,*}$ at $t=6.5 t_*$, while $\sigma$ in T1L is comparable to that in T3L, and T10L shows a lower $\sigma$ compared to T3L, while T0.01L gives the lowest $\sigma$. These results are consistent with Figure \ref{fig_dist}. The lower panel of Figure \ref{fig_sigmaL} shows that $\sigma$ is small compared to $V_{\rm mean}$, except for T3L which gives $\sigma \gtrsim 0.75 V_{\rm mean}$ after $t\gtrsim 3 t_*$.


\subsection{Small-Scale Runs}\label{section_smallscale}

As mentioned in Section \ref{section_cloudmodels}, the consideration to resolve the scale of gas turbulence motivates the small-scale runs. We carry out two runs T0.01S and T1S, which are compared to the large-scale runs T0.01L and T1L respectively. In small-scale runs we fix the column density of the cloud the same as that in the corresponding large-scale runs in cgs units, but shrink the cloud radii to $50 h_*$ with $h_*= c_{s,*}^{2}/g_*$. The ratio of the the initial average density of the cloud and the unit of time $t_*$ in small-scale (equation \ref{time_small}) and large-scale runs (equation \ref{time_large}) satisfies
\begin{equation}
\frac{\rho_*^{\rm S}}{\rho_*^{\rm L}}=\frac{t_*^{\rm L}}{t_*^{\rm S}}=\frac{h_*^{\rm L}}{h_*^{\rm S}}\simeq 89 T_{*,2}^{5}\label{SLratio},
\end{equation}
where $T_{*,2}=T_*/100\,$K. The density of cloud as well as the background in small-scale runs increase about two orders of magnitude compared to the large-scale runs. The dimensionless equations under code units for small-scale runs are almost the same as for the large-scale runs, expect for one single parameter $\Prat$, which measures the relative importance of radiation pressure over gas pressure (see Section \ref{section_reduced_speed1}). Mathematically the difference between large-scale and small-scale runs are caused by $\Prat$. A lower $\Prat$ gives a slower acceleration and less pressure on the cloud. We have $\Prat\simeq 8.9\times 10^{3}\tau_*^{-1}$ for large-scale runs and $\Prat\simeq 1.0\times 10^{2}\tau_*^{-1}$ for small-scale runs.  


Figure \ref{fig_SR} shows density snapshots from T0.01S (upper panels) and T1S (lower panels). We stop simulations at $t=100\, t_*$, which in cgs units is still shorter than $8\,t_*$ in large-scale runs, but the behavior of the cloud can be already well observed within $t\lesssim 100\,t_*$ in small-scale runs. The cloud in T0.01S is squeezed by radiation pressure and background pressure, and forms a weak pancake structure at $t\sim 75\,t_*$. However, since $\Prat$ is lower, the dimensionless density $\rho/\rho_*$ is also lower in T0.01S than that in T0.01L, and the cloud region near the background with lower $\rho/\rho_*$ is more likely to occur shearing instability. A Rayleigh-Taylor-like instability is developed at the front of the cloud interacting with the background, while the velocity difference between the cloud and the background at both sides of the cloud eventually shreds the cloud at $t\sim 90-100\,t_*$. On the other hand, in T1S a vertically elongated structure emerges by $t=50\,t_*$, while Kelvin-Helmholtz instability occurs at both sides of the cloud. In contrast to the filamentary structure in T1L, the interaction between the relatively low-density front of the elongated structure and the background fragments the front of the cloud into many small clumps. The cloud is eventually shredded by background pressure at $t\sim 100 \, t_*$ and fragments into pieces. 

Although the cloud morphology in small-scale runs is significantly different from that in large-scale runs, we find that the bulk acceleration of the cloud shows similar behavior. Figure \ref{fig_vmean_b} shows $V_{\rm mean}$ for T0.01S, T1S and T10S. We find that equations (\ref{analytic}) and (\ref{analytic2}) can also be applied for small-scale runs with $V_* \simeq c_{s,*}$, and the temperature of the cloud is still $T_{c}\sim 120\,$K, thus $V_{\rm mean}\sim1.4 V_* (t/t_*)$ also holds for T0.01S. Also, we obtain $\langle {\rm e}^{-\tau} \rangle \simeq 0.57$ for T1S, and $\langle {\rm e}^{-\tau} \rangle \simeq 0.14$ for T10S, which shows similar results compared to T1L and T10L respectively. Equation (\ref{taufitting2}) still roughly holds for cloud acceleration from $\tau_* =1$ to $\tau_*=10$. Changing the characteristic lengthscale of cloud may change the cloud morphology, but does not change cloud acceleration.

\begin{figure}[t]
\centerline{\includegraphics[width=8.5cm]{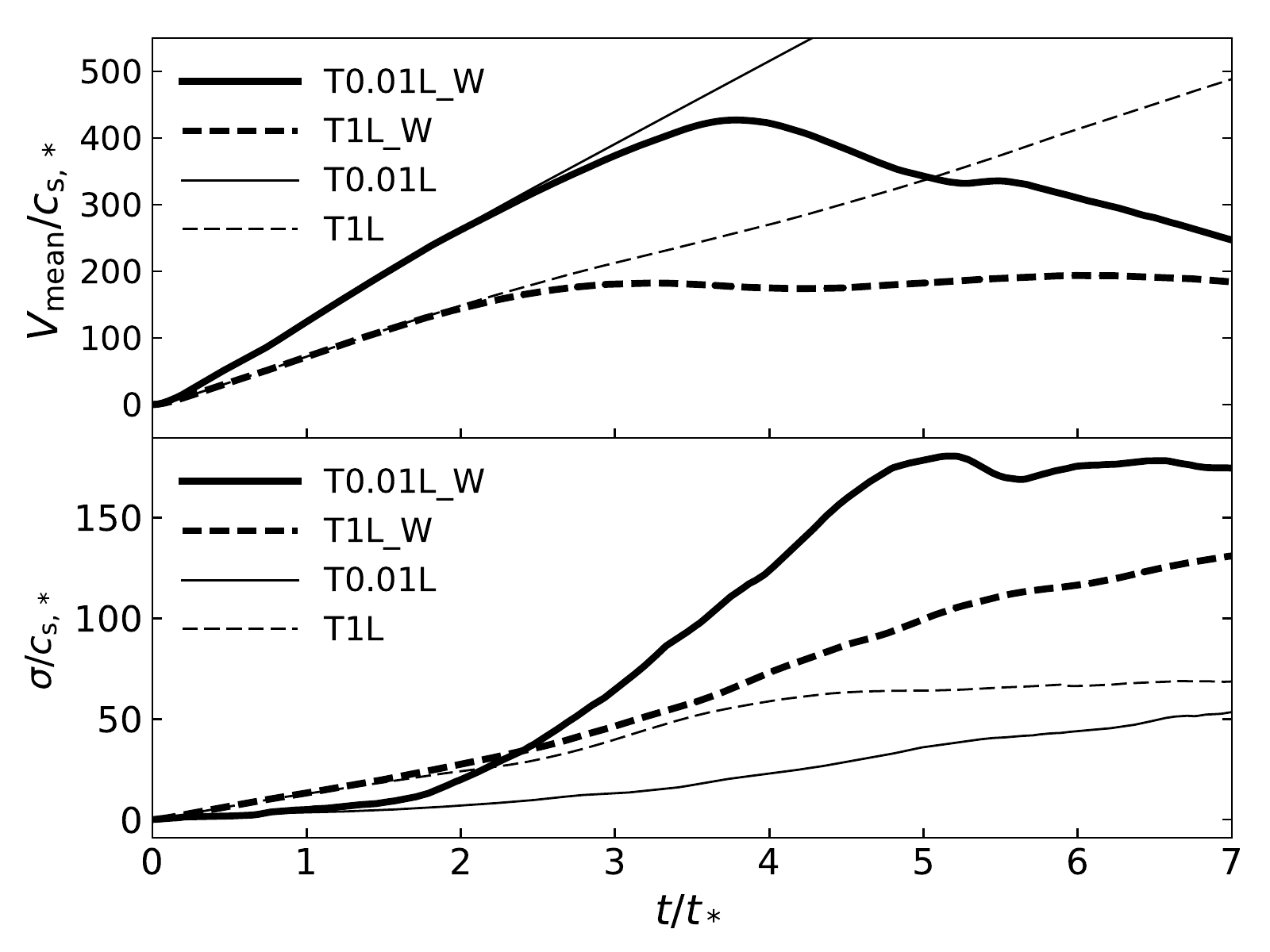}}
\caption{Cloud mean velocity $V_{\rm mean}$ (upper panel) and velocity dispersion $\sigma$ (lower panel) for the runs T0.01L\_W and T1L\_W compared to T0.01L and T1L. Here T0.01L\_W and T1L\_W are the large-scale runs with a warm background (W) setup $\chi_0=10^{2}$.}\label{fig_warmISM}
\end{figure}
\begin{figure}[t]
\centerline{\includegraphics[width=8cm]{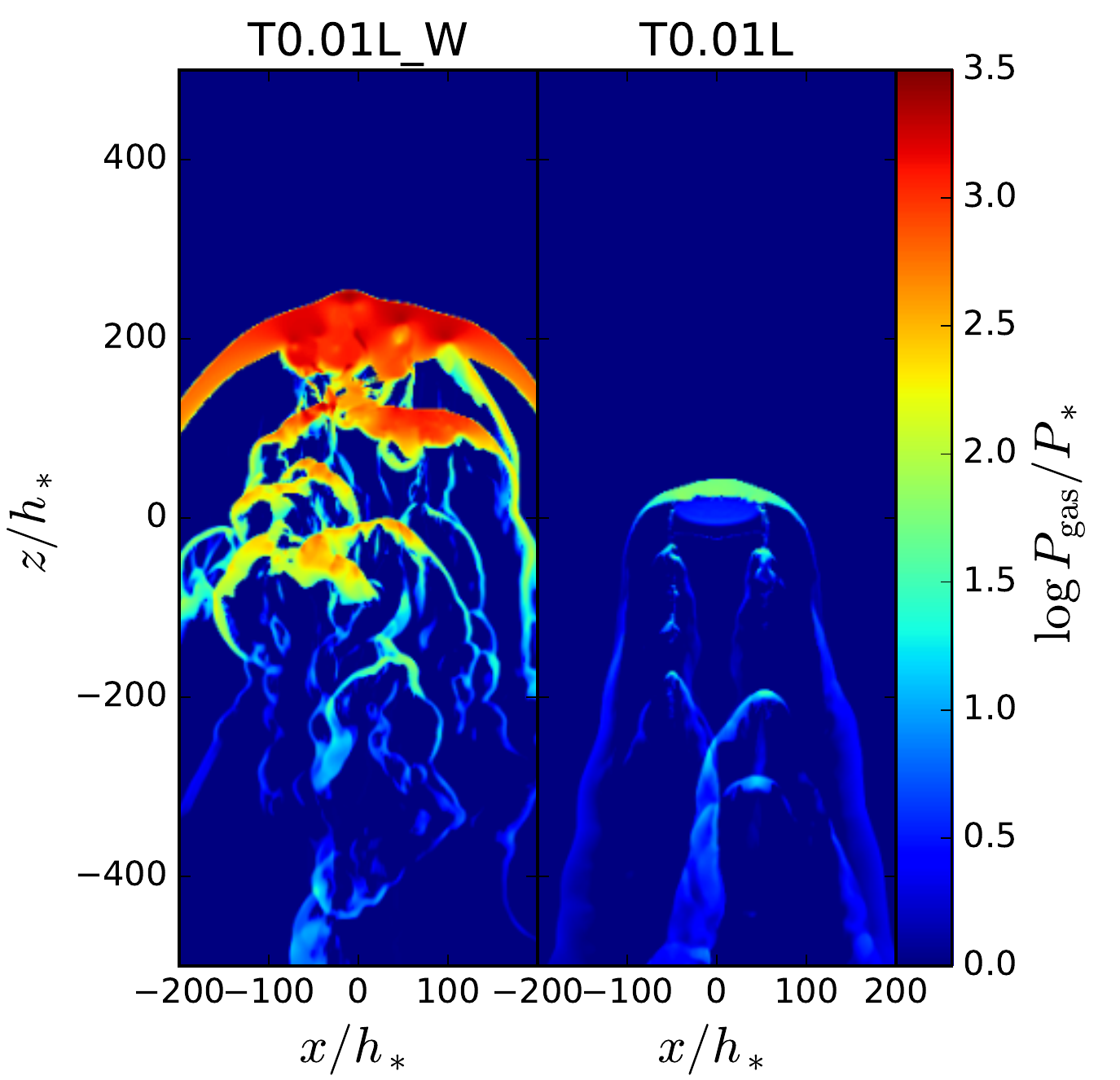}}
\caption{Comparison of thermal pressure in T0.01L\_W and T0.01L at $t=6\,t_*$. Here $P_* = \rho_* c_{s,*}^{2}$ is the unit of pressure and $P_{\rm gas}/P_*$ is the dimensionless pressure.}\label{fig_warmP}
\end{figure}
\begin{figure*}[t]
\centerline{\includegraphics[width=17cm]{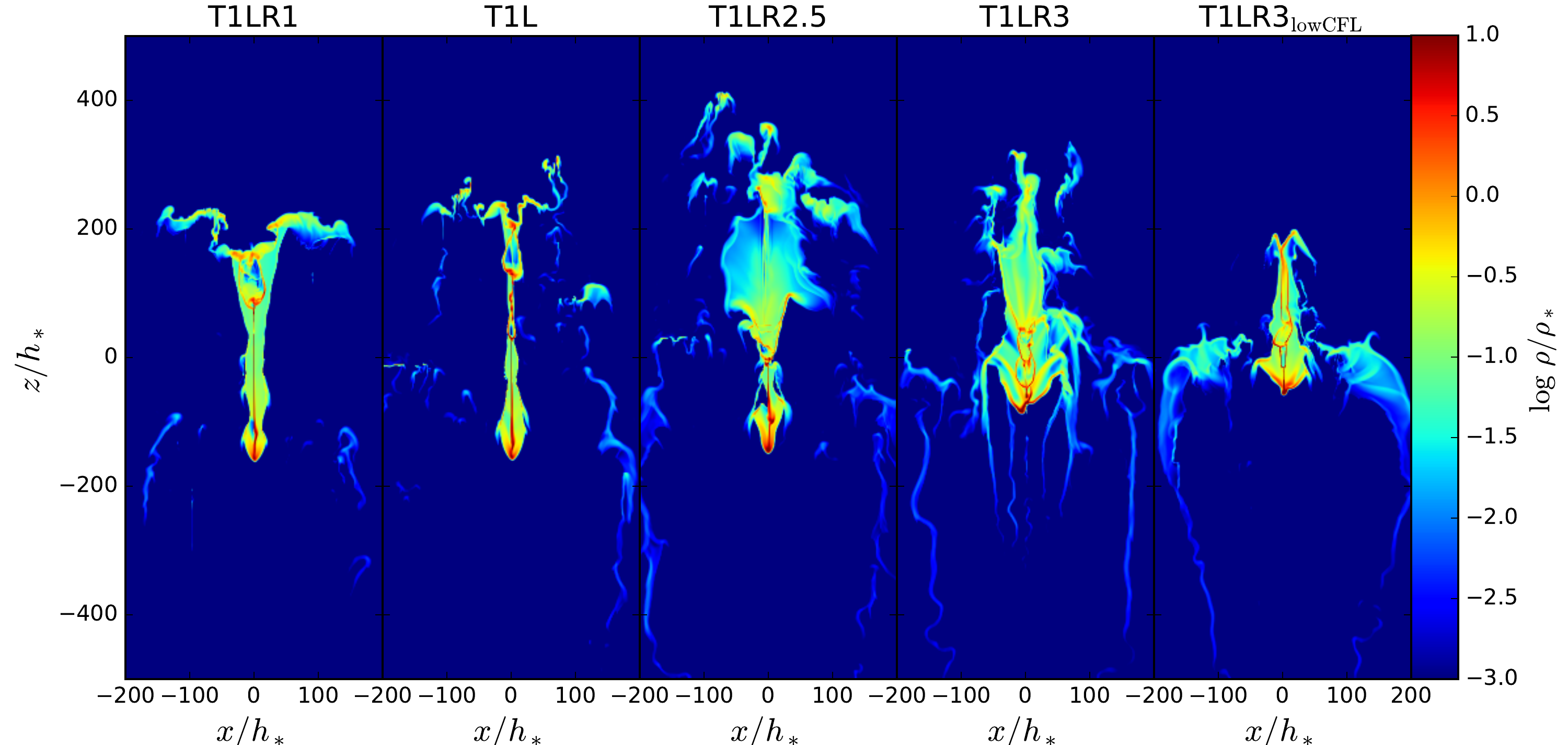}}
\caption{Comparison of density profiles from runs with different reduced speed of light at $t=5\,t_*$, where T1LR1, T1L, T1LR2.5 and T1LR3 correspond to large-scale runs with the reduction factor $\Rrat=10^{-1}$, $10^{-2}$, $10^{-2.5}$ and $10^{-3}$ respectively. The TLR3$_{\rm lowCFL}$ run has a lower CFL number so the timestep is exactly the same as in T1L.}\label{fig_reduced}
\end{figure*}
\begin{figure}[t]
\centerline{\includegraphics[width=9cm]{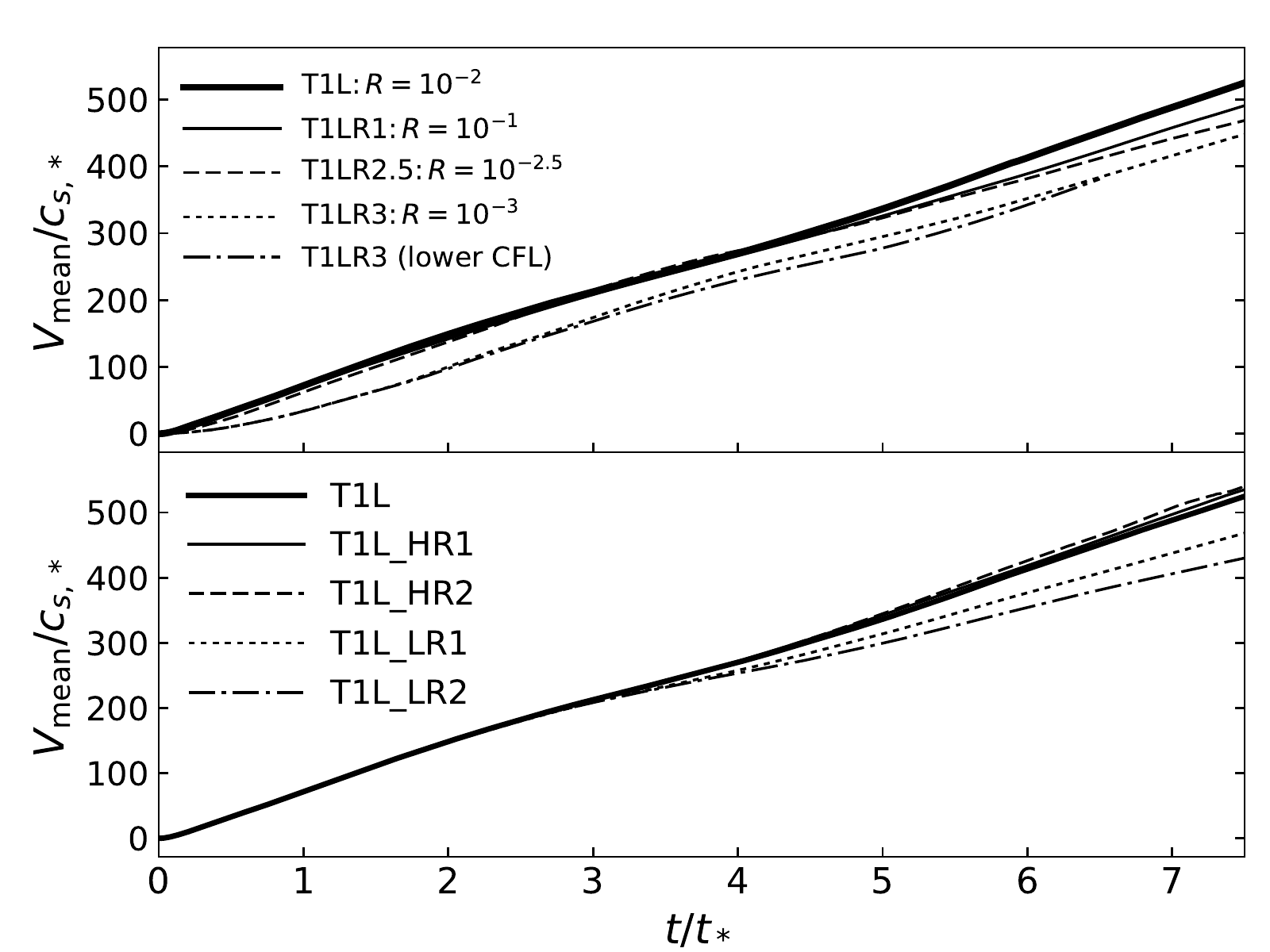}}
\caption{Cloud mean velocities for different reduced speed of light T1L, T1LR1, T1LR2.5 and T1LR3 (upper panel), and different spatial resolution T1L, T1L\_HR1, T1L\_HR2, T1L\_LR1 and T1L\_LR2 (lower panel), where the definitions of ``HR1", ``HR2", ``LR1" and ``LR2" are given in Table \ref{tab_parameter}.}\label{fig_red_res}
\end{figure} 
\begin{figure*}[t]
\centerline{\includegraphics[width=17cm]{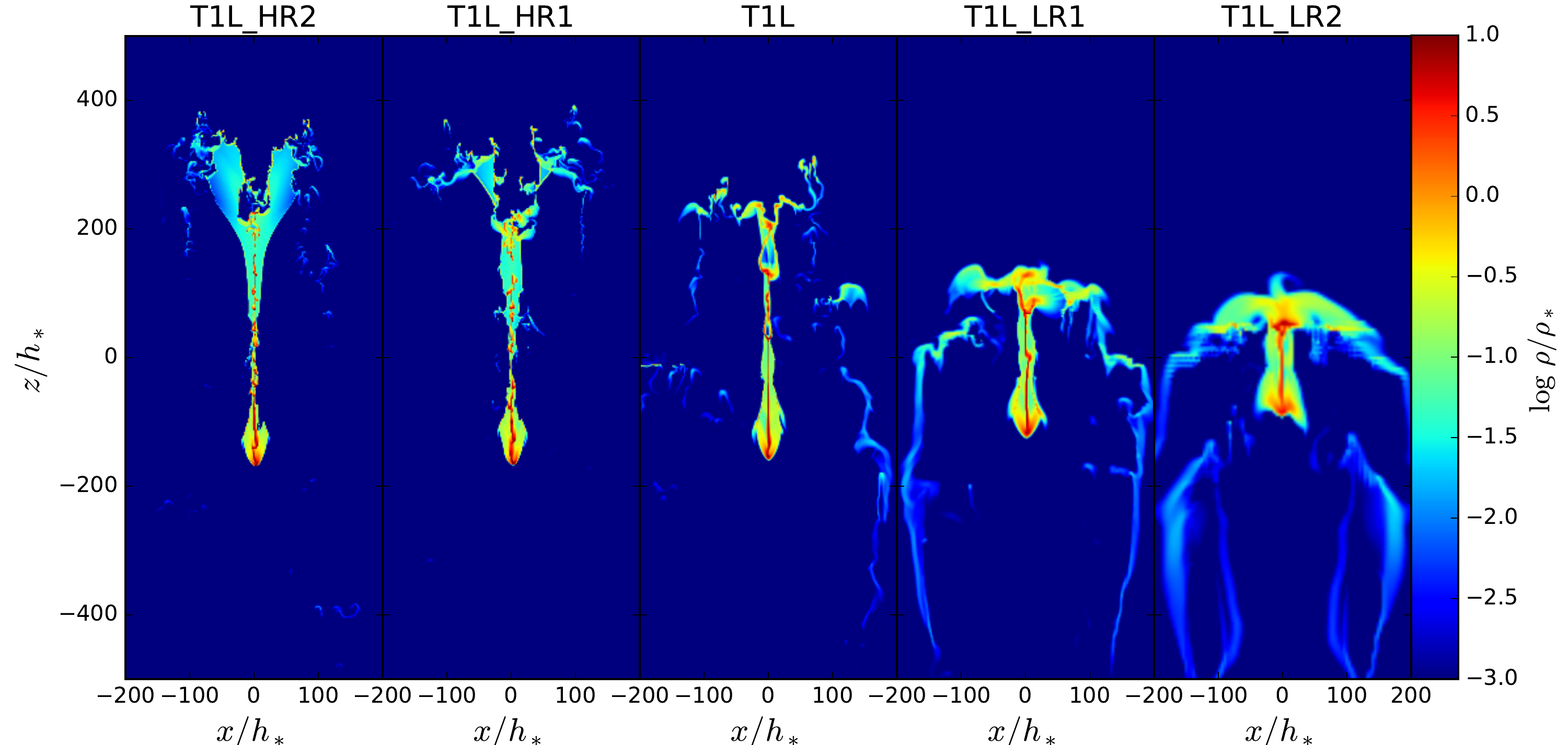}}
\caption{Comparison of runs for various spatial resolutions T1L\_HR2, T1L\_HR1, T1L, T1L\_LR1, T1L\_LR2 at $t=5t_*$.}\label{fig_resolution}
\end{figure*}

\subsection{Warm Background Medium}\label{section_warmISM}

So far we discuss clouds in a hot background medium with density ratio $\chi_0 =10^{4}$. Here we consider clouds in a warm background where has a temperature of $T_{\rm bkgd}=10^{4}\,$K and $\chi_0=T_{\rm bkgd}/T_*=10^{2}$. Since thermal pressure and ram pressure from the warm background are significantly stronger than those from the hot background, the interaction between the cloud and the background may cause different behavior in contrast to the cloud in a hot background. We carry out two runs T0.01L\_W and T1L\_W to study the impact of denser background. Figure \ref{fig_warmISM} shows $V_{\rm mean}$ and $\sigma$ for these two runs which are compared to T0.01L and T1L respectively. In contrast to T0.01L and T1L, the mean velocity of the cloud in T0.01L\_W increases to $V_{\rm mean}\sim 420\,c_{s,*}$ at $t\simeq 3.8 t_*$, then the cloud begins to be decelerated, while $V_{\rm mean}$ in T0.01L\_W stays flat at $V_{\rm mean}\sim 180-190\,c_{s,*}$ for $t\gtrsim 3\,t_*$. The lower panel in Figure \ref{fig_warmISM} shows that the velocity dispersions in T0.01\_W and T1\_W are also significantly larger than T0.01L and T1 respectively. In particular, the velocity dispersion is comparable to mean velocity $V_{\rm mean}$ at $t\sim 7 t_*$ in T0.01L\_W, because the cloud becomes very turbulent and is eventually mixed up with the background.

We find that thermal pressure in the shock-like interface between the front of the cloud and background dominates over ram pressure from the background. In other words, thermal pressure plays more important role to re-shaping and decelerating the cloud. Figure \ref{fig_warmP} shows thermal pressure $P_{\rm gas}/P_*$ in T0.01L\_W and T0.01L at $t=6\,t_*$. Note that the high pressure region in Figure \ref{fig_warmP} is not in the cloud, but in the interface region between the cloud and non-disrupted background. We find that the temperature in the shocked interface between the cloud and background is $T\sim 10^{4}T_*$ in T0.01L\_W, while $T\sim 10^{5}T_*$ in T0.01L, thus the interface gas pressure reaches $P_{\rm gas}\sim 10^{3}P_*$ in T0.01L\_W, while $P_{\rm gas}\sim 50 P_*$ in T0.01L, which is one and half order of magnitude lower compared to the warm background. The high gas pressure acting on the cloud from the interface eventually prevents the cloud from being accelerated by radiation, and shreds the cloud more quickly than the cloud in a hotter and more tenuous background.

Another important quantity to measure the cloud properties is the cloud survival time, which is investigated in Section \ref{section_hotwind2}. Also cooling and heating in the ionized background can change the cloud dynamics and morphologies, which is discussed in Section \ref{section_cooling}. Before discussing them we study the effects of some simulation parameters such as reduced speed of light, resolution and simulation dimensionality in the following several sections.

\subsection{Reduced Speed of Light Approximation}\label{section_reduced_speed2}

The details of the reduced speed of light approximation are discussed in Appendix \ref{append_reduced_speed}. Clouds in large-scale runs has a maximum velocity $v_{\rm max}\sim 500-10^{3} c_{s,*}$ and optical depth $\tau_{\rm max}\sim 1$, thus we have the constraint $500-1000\ll \tilde\Crat\ll \Crat$ using the criterion equation (\ref{skinner}) in Appendix \ref{append_reduced_speed}. The lower bound of the reduction factor is $\Rrat= \Crat/\tilde\Crat \gg 1.5-3 \times 10^{-3}$. Therefore it is justified to use $\Rrat=10^{-2}$ as a fiducial value in the paper. We take T1L as a fiducial run and perform other three runs with various reduction factor $\Rrat$ that T1LR1 ($\Rrat=10^{-1}$), T1LR2.5 ($\Rrat=10^{-2.5}$) and T1LR3 ($\Rrat=10^{-3}$) to investigate the effects of $\Rrat$. The left four panels of Figure \ref{fig_reduced} compares the cloud density profiles for varying $\Rrat$ at $t=5\,t_*$. The cloud morphologies are similar in T1LR1 and T1L, while the filamentary structure slightly changes in T1LR2.5 and becomes less vertically extended in T1LR3. We also find that velocity dispersion varies for different $\Rrat$, which result is consistent with Figure \ref{fig_reduced} that the shape and properties of turbulence depends on $\Rrat$. However, most of the cloud mass is still accumulated in the lower region of the cloud along the vertical direction, and cloud acceleration which mainly depends on the bottom region of the cloud does not change significantly. The upper panel of Figure \ref{fig_red_res} shows $V_{\rm mean}$ for these runs. We find that $V_{\rm mean}$ does not change obviously for $\Rrat\geq 10^{-2.5}$. Even $\Rrat=10^{-3}$ gives less than $\sim 20\%$ difference in $V_{\rm mean}$. Thus we conclude that $10^{-3} \lesssim \Rrat \lesssim 1$ gives a good approximation for cloud acceleration.

A higher $\Rrat$ gives a smaller timestep in simulations. So far in all 2D simulations we choose the Courant--Friedrichs--Lewy (CFL) number to be 0.4 (see Appendix \ref{append_reduced_speed}), it is worthwhile to test whether it is equivalent to use a lower CFL number instead of a higher $\Rrat$ to decrease the timestep, or $\Rrat$ is more important to affect the RT. We perform another T1LR3 run (T1LR3$_{\rm lowCFL}$) with $\Rrat=10^{-3}$ but has a smaller CFL number so the timestep is exactly the same as in T1L. The rightmost panel of Figure \ref{fig_reduced} shows that the geometry of the cloud in T1LR3$_{\rm lowCFL}$ is more like that in T1LR3 rather than in T1L, and the upper panel of Figure \ref{fig_red_res} shows that the cloud velocity for T1LR3$_{\rm lowCFL}$ is almost identical to that for T1LR3. From the aspect of cloud evolution, we conclude that varying $\Rrat$ is more important to affect the properties of RT equation rather than to vary timestep in hydrodynamic equations.

We also test the reduction factor in small-scale runs. Note that $v_{\rm max}$ is much lower in small-scale runs, equation (\ref{skinner}) gives a much lower estimate on the lower bound of $\Rrat$ that $\Rrat \gg 1.5 \times 10^{-4}$ for $v_{\rm max}\sim 100\,c_{\rm s,*}$.  However, we find that even $\Rrat=10^{-4}$ gives an good approximation of both $V_{\rm mean}$ and velocity dispersion $\sigma$ compared to higher $\Prat$. We report that the small-scale runs allow a much weaker constraint on $\Prat$.

\subsection{Dependence on Spatial Resolution}\label{section_resolution}

We also study the impact of spatial resolution on cloud simulations. The fiducial run T1L assumes that $\Delta z/h_* =1$ (see Table \ref{tab_parameter}). Compared to the fiducial run, a number of other runs are also performed here, with higher resolutions $\Delta z/h_*=0.5$ (T1L\_HR1), $\Delta z/h_*=0.25$ (T1L\_HR2), and lower resolutions $\Delta z/h_*=2$ (T1L\_LR1) and $\Delta z/h_*=4$ (T1L\_LR2).

Figure \ref{fig_resolution} compares density snapshots from five runs with various resolutions at $t=5\,t_*$. The shapes of the cloud are obviously different in the low-resolution runs from the high-resolution runs. In general, higher spatial resolution run gives a longer stretched cloud along the vertical direction. More specially, we find that the resolution in $z$-direction controls the behavior of the cloud. Same resolution in $x$-direction but higher resolution in $z$-direction (e.g., T1L\_HR2 compared to T1L\_HR1) gives a more rapidly evolved cloud with higher turbulence, while we also test that different $x$-direction but same $z-$direction resolution shows similar cloud evolution. A lower $z$-direction resolution makes cloud more compact along the vertical direction. In spite of these differences, the filamentary structure still appears in all the runs, with most mass distributed at the bottom of the clouds. The lower panel of Figure \ref{fig_red_res} shows $V_{\rm mean}$ with various resolution runs. The cloud bulk acceleration $V_{\rm mean}$ is almost identical from the highest resolution run T1L\_HR2 to the moderate resolution run T1L, which shows that T1L is justified as a fiducial resolution. Lower resolution runs show slower acceleration,  but the difference is within $\sim 20\%$ between T1L and the lowest-resolution run T1L\_LR2, which means that $V_{\rm mean}$ only depends weakly on spatial resolution.  

On the other hand, we find that the impact of spatial resolution is much less important for optically thin cloud. We carry out two large-scale runs T0.01L\_HR1, T0.01L\_LR1 as shown in Table \ref{tab_parameter}) and compare to T0.01L, but find almost no difference of $V_{\rm mean}$ among them. Since the acceleration depends on $\kappa F/c$, which is almost uniformly distributed inside the cloud both for lower and higher-resolution runs, the evolution of the cloud is very similar in runs with different resolutions.

\begin{figure}[t]
\centerline{\includegraphics[width=8.5cm]{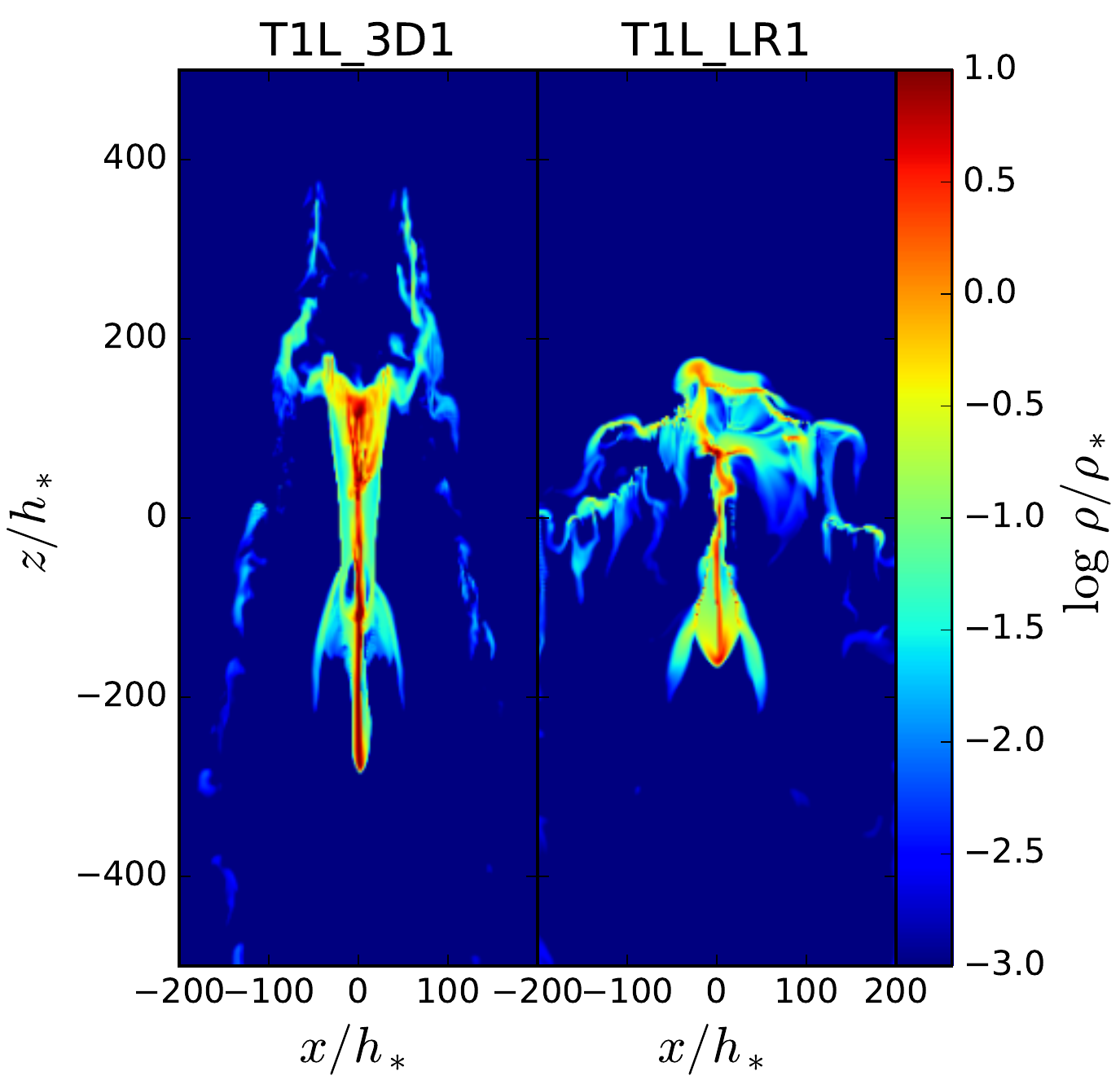}}
\caption{Comparison of density profiles from 3D run T1L\_3D1 and 2D run T1L\_LR2 at $t=6\,t_*$. The 2D $xz$ slice with $y=0$ is plotted for the 3D panel.}\label{fig_3D1}
\end{figure}
\begin{figure}[t]
\centerline{\includegraphics[width=9.5cm]{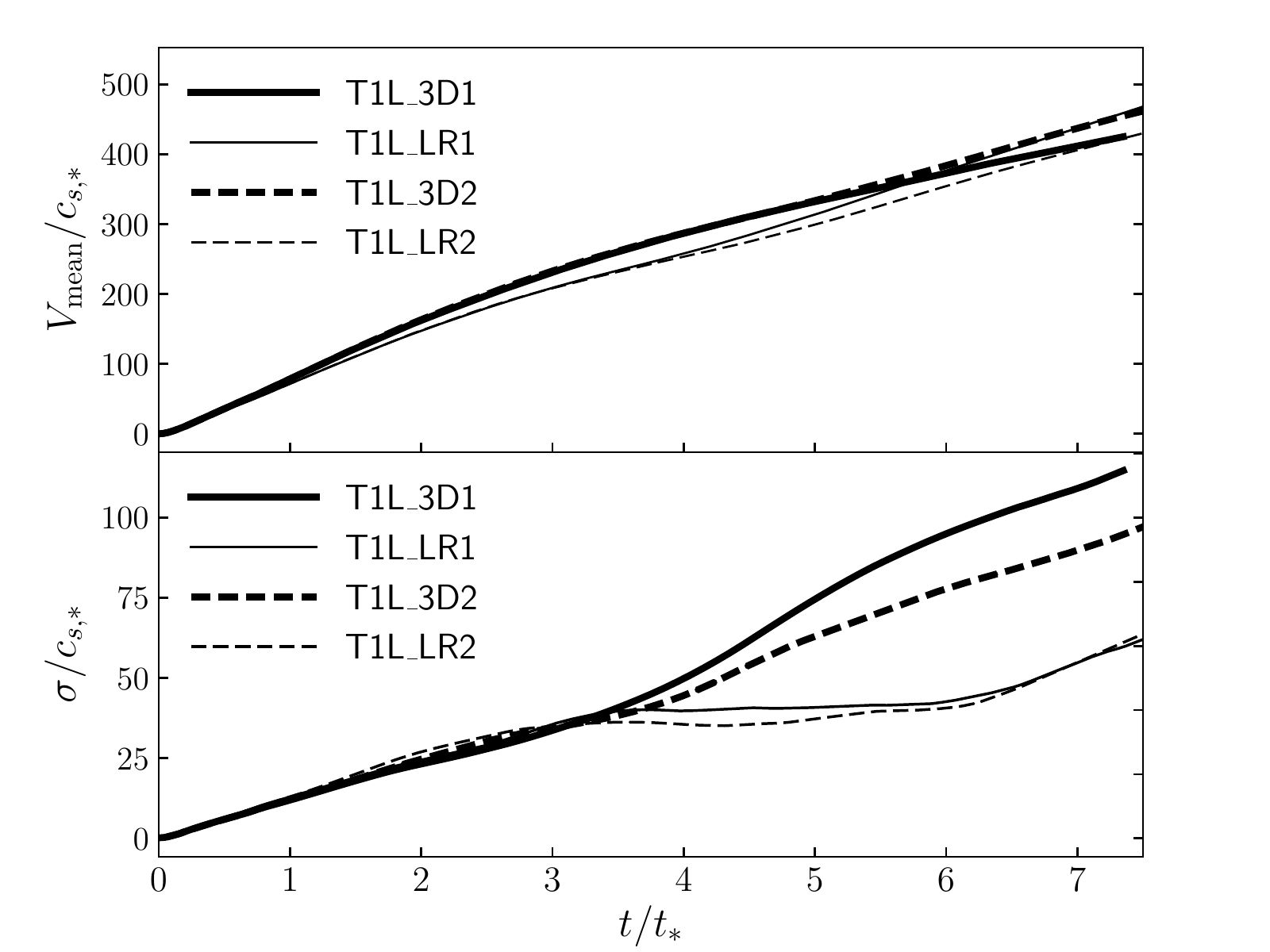}}
\caption{Cloud mean velocities (upper panel) and velocity dispersion (lower panel) for the 3D runs T1L\_3D1 and T1L\_3D2, and the equivalent 2D runs T1L\_LR1 and T1L\_LR2.}\label{fig_3D2}
\end{figure}
\begin{figure*}[t]
\centerline{\includegraphics[width=17cm]{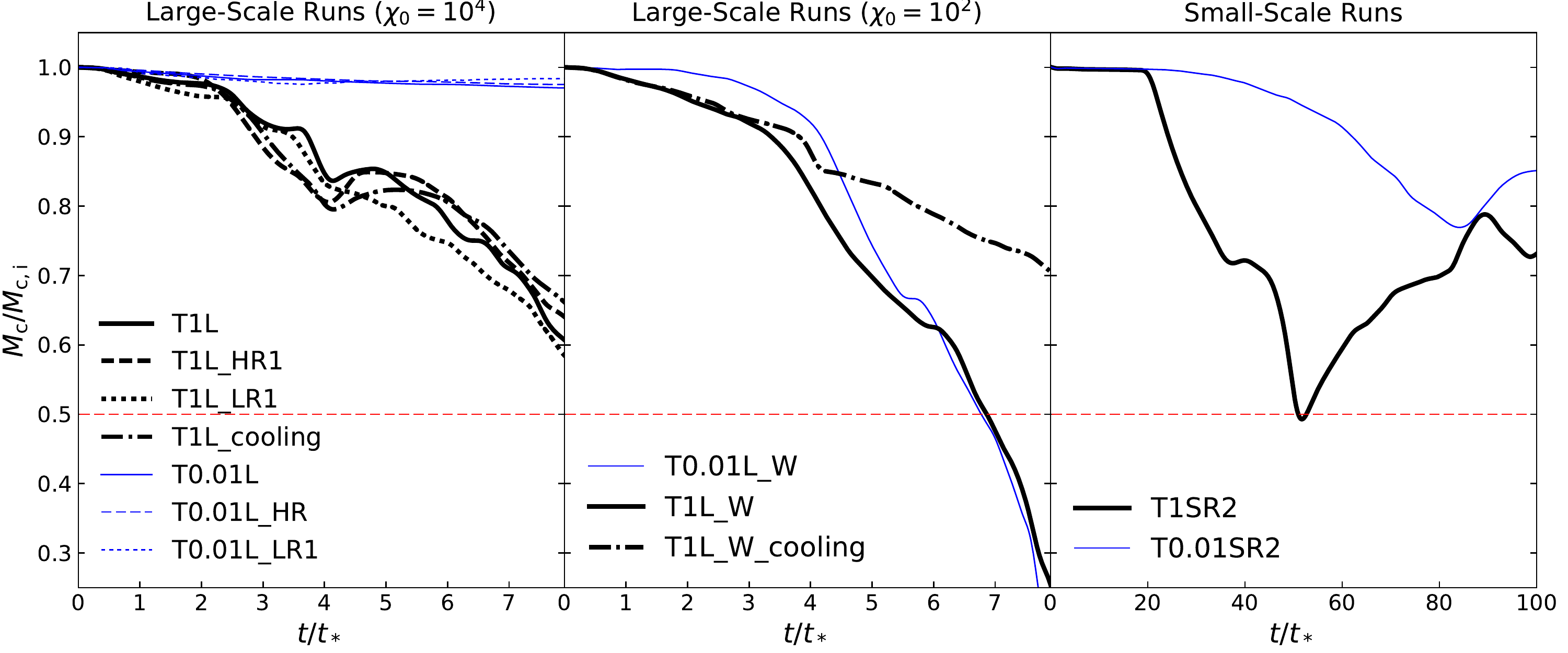}}
\caption{Evolution of cloud mass, which is calculated as the sum of all gas in the simulation with a density greater than $1/3$ the initial median density, for large-scale runs in hot background ($\chi_0 =10^{4}$, left panel), warm background ($\chi_0 =10^{2}$, middle panel), and small-scale runs (right panel). The runs in left panel are from Sections \ref{section_thin}, \ref{section_thick}, \ref{section_resolution}, runs in middle panel are from Section \ref{section_warmISM}, and runs in right panel are from \ref{section_smallscale}.}\label{fig_mass}
\end{figure*}
\begin{figure}[t]
\centerline{\includegraphics[width=9cm]{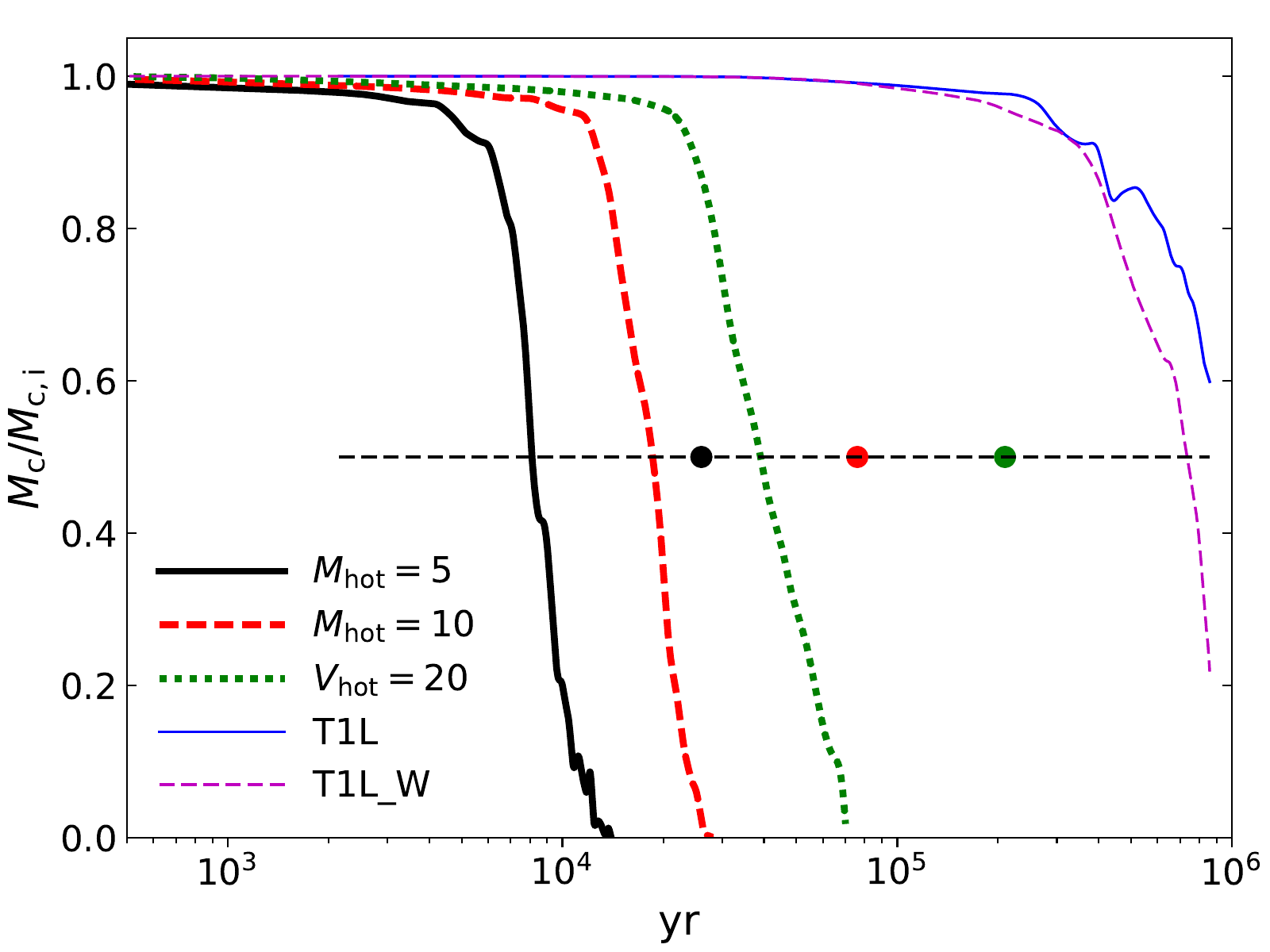}}
\caption{Mass evolution of cloud accelerated by hot wind runs H1, H2 and H3 with hot flow Mach number $M_{\rm hot}=5,10,20$ respectively, and radiatively accelerated runs T1L and T1L\_W for comparison.  Three dots from left to right are the values of $t_{50}$ given by equation (\ref{survivaltime}) for $M_{\rm hot}=5,10,20$ respectively.}\label{fig_hotcloud}
\end{figure}

\subsection{3D Simulations}\label{section_3D}

So far we discuss the simulation results based on 2D simulations, but 3D simulation may change some properties of the cloud. Optically thick cloud develops non-linear structure which may be different in 2D and 3D simulations. Since 3D runs are more expensive than 2D runs, we only carry out two runs: T1L\_3D1 and T1LR2\_3D2, which are equivalent to T1L\_LR1 and T1L\_LR2 runs in 2D with the same spatial resolution respectively. We set the CFL number to be 0.3 in 3D runs, thus the time-steps in 3D runs are relatively smaller compared to the 2D runs with the same spatial resolution. In Section \ref{section_resolution} we show that although the cloud morphologies vary for different resolution, the cloud bulk acceleration is still similar. Therefore it is justified to adopt low-resolution 2D runs and see the difference of cloud evolution with an extra dimension.

Figure \ref{fig_3D1} compares density distribution in T1L\_3D1 with T1L\_LR1 at a same time $t=6\,t_*$. Two runs show a bit different cloud evolution. The behavior of the cloud is more like T1L rather than T1L\_LR1. The 3D cloud is stretched out and elongated along the vertical direction from $z\sim -300\,h_*$ to $200\,h_*$, which shape is similar to T1L and other 2D higher resolution runs. Also the filamentary structure in T1L\_3D1 has a typical value of $\sim 5-10 \rho_*$, which is slightly denser compared to the cloud in T1L\_LR1, but is comparable to T1L. In spite of the difference, we find that the values of $V_{\rm mean}$ do not change too much in these runs. The upper panel of Figure \ref{fig_3D2} shows that the difference of $V_{\rm mean}$ between 2D and 3D runs is less than $\sim 10\%$. The lower panel of Figure \ref{fig_3D2} shows that velocity dispersion of the cloud can significantly increase to $\sigma\sim 120 c_{s,*}$, which means that 3D cloud is more turbulent than 2D cloud. The comparison suggests that 2D runs are good approximations to model cloud acceleration, although the detailed geometry and turbulence of the clouds depend on the dimensionality of runs.

\section{Discussion}\label{section_discussion}

\subsection{Cloud Survival Time}\label{section_hotwind2}

We follow \cite{SB15} and \cite{SR17} to define the cloud mass as the summation of all gas denser than $1/3$ of the initial cloud mean density $\rho_*$. Figure \ref{fig_mass} shows the cloud mass evolution, i.e., the ratio of the cloud mass to the initial cloud mass $M_{c}/M_{c,\textrm{i}}$ as a function of time, for large-scale runs in the hot  ($\chi_0=10^{4}$) and warm ($\chi_0=10^{2}$) background, and for small-scale runs. In large-scale runs an optically thin cloud in the hot background ($\tau_*=0.01$ and $\chi_0=10^{4}$) shows the most ideal acceleration, in which the mass of cloud is almost unchanged during its acceleration. In the warm background a same optically thin cloud is decelerated by thermal pressure of the background at $t\sim 3.8\,t_*$ (see Figure \ref{fig_warmISM}), meanwhile the cloud begins to be shredded, and $M_{c}/M_{c,\textrm{i}}$ quickly drops from $M_{c}/M_{c,\textrm{i}} \sim 0.9$ to $\sim 0.5$ at $t\sim7\,t_*$. For large-scale runs with $\tau_*=1$, about $\sim 70\%-75\%$ of the initial mass is still in the cloud at $t=7\,t_*$, while the mass fraction $M_{c}/M_{c,\textrm{i}}$ drops to $\sim 50\%$ in the case of warm background, slightly lower than that in a hot background, but similar to the optically think cloud in the warm background. On the other hand, the evolution of cloud mass in small-scale runs show an interesting behavior. The cloud mass fraction drops to $M_{c}/M_{c,\textrm{i}} \sim 0.5$ at $t\sim 50\,t_*$ in T1SR2, but the mass fraction increases again afterwards. Figure \ref{fig_SR} and the right panel of Figure \ref{fig_mass} shows that the cloud is initially stretched and the cloud density drops until $\sim 50\,t_*$ when the cloud starts to fragments into pieces. The density of many fragmentations increases again up to $\rho\sim \rho_*$ because the pressure of the background confines and squeezes these small pieces. Then $M_{c}/M_{c,\textrm{i}}$ rises up again to $M_{c}/M_{c,\textrm{i}}\sim 0.8$. A similar behavior also emerges in T0.01S, in which the mass of the cloud drops to $M_{c}/M_{c,\textrm{i}}\sim 0.7$ at $t\sim 80\,t_*$ and increases again afterwards. We do not find significant decrease of $M_{c}/M_{c,\rm i}$ below 0.5 for $t\lesssim 100 t_*$.

If we define the cloud survival time as the time when $M_{c}/M_{c,\textrm{i}}$ drops to 0.5, and denote it as $t_{50}$ (see \citealt{SB15}), we find that $t_{50}$ mainly depends on $\tau_*$ and $\chi_0$. The cloud survival time $t_{50}$ is longer than the simulation time in runs with $\chi_0 =10^{4}$ both for large-scale and small-scale runs, while the cloud survival time is significantly shorter in the warm background $\chi_0 =10^{2}$ that $t_{50} \sim 7\, t_*$ in both T0.01L\_W and T1L\_W. Nevertheless, we find that the lifetime of the cloud is still significantly longer than that in a hot wind.  
 
In order to compare cloud in radiation and in hot wind, we perform three hydrodynamic simulations H1, H2 and H3 with Mach number of the hot wind $M_{\rm hot}=5,10,20$ (see Section \ref{section_hotwind1} and Table \ref{tab_parameter}). We translate code units back to cgs units for comparisons. The evolution of cloud mass in hot winds $M_{c}/M_{c,\rm i}$ is shown in Figure \ref{fig_hotcloud}. Compared to T1L and T1L\_W runs, we find that cloud mass drops significantly faster in hot winds. According to our simulations, the cloud survival time $t_{50}\sim 8.6\times 10^{3}\,$yr for H1 ($M_{\rm hot}=5$), $t_{50}\sim 1.8\times 10^{4}\,$yr for H2 ($M_{\rm hot}=10$), and $t_{50}\sim 3.6\times 10^{4}\,$yr for H3 ($M_{\rm hot}=20$) respectively. The values of $t_{50}$ we obtained are relatively lower than those in \cite{SB15} (see also \citealt{SR17}), who gives $t_{50}=\alpha \sqrt{1+M_{\rm hot}}t_{\rm cc}$ with the crushing time if the cloud in a hot flow $t_{\rm cc}=\chi_0^{1/2}(R_{c}/V_{\rm hot})$. \cite{SB15} found that $\alpha \approx 4$, but we find lower efficiency that $\alpha\approx 1.3$, 0.9 and 0.7 in H1, H2 and H3 respectively. The differences can be caused by the adiabatic background and the parameter sets we used in the paper. We choose the initial cloud density $\rho_* \sim 10^{-20}\,$g cm$^{-3}$, which is four orders of magnitude higher than the cloud density in \cite{SB15}; and we use $T_{\rm hot}=10^{6}\,$K and $T_{c}=10^{2}\,$K, which are lower than the values in \cite{SB15}. However, even we adopt $t_{50}$ in \cite{SB15} for our parameter set, we have 
\begin{equation}
t_{50}^{\rm SB}\simeq 2.4 \times 10^{3}\,\textrm{yr}\,(\alpha/4)M_{\rm hot}^{3/2}\tau_*\label{survivaltime},
\end{equation}
thus in H1, H2 and H3 the clouds have $t_{50}^{\rm SB}\simeq 2.6\times 10^{4}\,$yr, $t_{50}^{\rm SB}\simeq 7.6\times 10^{4}\,$yr and $t_{50}^{\rm SB}\simeq 2.1\times 10^{5}\,$yr for $M_{\rm hot}=5$, 10 and 20 respectively (see the dots in Figure \ref{fig_hotcloud}), which are still significantly lower than $t_{50}$ obtained from the cloud radiation runs $t_{50}\gg 10^{6}\,$yr for T1L and $t_{50}\sim 7.5\times 10^{5}\,$yr for T1L\_W. Equation (\ref{survivaltime}) also shows that the cloud survival time is independent of $\chi_0$. We conclude that the survival time of the cloud $t_{50}$ is much longer than $t_{50}$ of the cloud entrained in a hot wind. The survival time of clouds in various runs are summarized in Table \ref{tab_survivaltime}.

\begin{table}[t]
\begin{center} 
Cloud Survival Time Results

\begin{tabular} {lcllccllc}
\hline\hline
Runs & $t_{50}$ (yr) & $t_{50}^{\rm SB}$ (yr) \\
\hline
T0.01L\_W &  $7.3\times 10^{5}$ &   \\
T1L\_W    &  $7.4\times 10^{5}$ &      \\
T1L    &  $> 10^{6}$ &      \\
$M_{\rm hot}=5$    &  $8.6 \times 10^{3}$ &  $2.6 \times 10^{4}$    \\
$M_{\rm hot}=10$    &  $1.8 \times 10^{4}$ &  $7.6 \times 10^{4}$    \\
$M_{\rm hot}=20$    &  $3.6 \times 10^{4}$ &  $2.1 \times 10^{5}$    \\
\hline
\end{tabular}
\end{center}
\caption{Note: The cloud survival time $t_{50}$ is obtained from our simulations mentioned in Section \ref{section_hotwind2}. The hot wind runs H1, H2 and H3 have hot flow Mach number $M_{\rm hot}=5,10,20$ respectively, while $t_{50}^{\rm SB}$ is obtained from \cite{SB15} (equation \ref{survivaltime} in this paper) with the same hot flow Mach numbers. }\label{tab_survivaltime}
\end{table}

\subsection{Cooling and Heating in the Background}\label{section_cooling}

So far we assume the background is adiabatic. Once the radiation pressure accelerates the cloud, the interface between the cloud and background is heated and becomes hotter than the background. The cooling timescale in the background and the interface region can be estimated by $t_{\rm cool}\sim 3kT/(n_{\rm hot}\Lambda(T))$, where $n_{\rm hot}$ is the density of the hot gas, and the cooling function $\Lambda(T)$ can be approximated as (\citealt{Draine11})
\begin{equation}
\Lambda(T) \simeq 1.1\times10^{-22}\,T_{6}^{-0.7}\,\textrm{erg}\,\textrm{cm}^{3}\,\textrm{s}^{-1}
\end{equation}
for $10^{5}\,$K$\lesssim T \lesssim 10^{7.3}\,$K, where $T_{6}=T/10^{6}\,$K is the gas temperature in the interface region. Thus, we estimate the cooling timescale for large-scale runs is
\begin{equation}
t_{\rm cool}\sim 9.8\times 10^{4}\,\textrm{yr}\,T_{6}^{1.7}\chi_{0,4}\tau_*^{-1}.
\end{equation}
Compared to equation (\ref{time_large}), we find that the condition $t_{\rm cool}\lesssim t_*$ holds for $\chi_0 =10^{4}$ with $\tau_* = 1$. The hot background with $T_{\rm bkgd} = 10^{6}\,$K can quickly cool down before the cloud is accelerated. However, heating either from supernova explosions or superwinds from massive stars, or photoheating in the rapidly star-forming environment can balance cooling in the hot background so that the background remains hot. 

Here we carry out several runs to study the effects of cooling and heating in the background. T1L\_cooling is based on the fiducial run T1L but includes cooling and heating in the background, other initial conditions are the same as in T1L. We compute the cooling rate from interpolating the cooling table in \cite{Schure09}, who calculated the cooling rate of a hot plasma in collisional ionization equilibrium (see also \citealt{Wiersma09}, \citealt{OS13}). The heating rate $\Gamma$ is assumed to balance the initial cooling rate in the background so the un-disrupted background keeps a constant temperature $T_{\rm bkgd} = 10^{6}\,$K. We find that heating and cooling in the background does not affect dust temperature in the cloud, therefore $V_{\rm mean}$ is nearly identical compared to that in T1L. However, the interface between the cloud and background cools down to $\simeq T_{\rm bkgd}$ thus the pressure from the interface region is lower than in T1L, and the filamentary structure along the vertical direction expands faster compared to that in T1L. The change of cloud morphology may also change the turbulence properties. However, the left panel of Figure \ref{fig_mass} shows that cooling does not change the evolution of $M_{c}/M_{c,\rm i}$ dramatically in the hot background. 

On the other hand, although CIE cooling is insignificant in the warm background with $T_{\rm bkgd}=10^{4}\,$K, it may still play important role to cool the interface between the cloud and background and to confine the shape of the clouds. We carry out two runs T0.01L\_W\_cooling and T1L\_W\_cooling which are based on T0.01L\_W and T1L\_W respectively, but include CIE cooling in the background. We find that the interface between the cloud and background cools down so that the clouds feel less thermal pressure from the interface, the asymptotic cloud mean velocity for T1L\_W\_cooling is $V_{\rm mean}\sim 330 c_{s,*}$, slightly higher than $V_{\rm mean}\sim 200 c_{s,*}$ in Figure \ref{fig_warmISM}. Also the cloud can obtain a longer survival time. The middle panel of Figure \ref{fig_mass} shows that the cloud mass evolution T1L\_W\_cooling drops more slowly than the run without cooling. However, the longer survival time of the irradiated cloud make the conclusion in Section \ref{section_hotwind2} even more robust. Moreover, the $V_{\rm mean}$ profile and $M_{c}/M_{c, \rm i}$ for T0.01L\_W\_cooling do not change significantly compared to the run with an adiabatic background.

Also, adding cooling and heating in the hot background to maintain $T_{\rm bkgd}=10^{6}\,$K may change the cloud evolution in hot flows. We find that $t_{\rm 50}$ increases by a factor of several compared to the values of H1, H2 and H3 runs. However, they are still shorter than the constraint by equation (\ref{survivaltime}). As a result, the survival time of the entrained clouds in hot flows are still shorter than it in a radiation field. In summary, cooling and heating in both hot and warm ionized background may slightly change the cloud morphologies, but do not change our main results qualitatively.

\subsection{Cloud Turbulence and Molecular Weight}\label{section_secondfactors1}

The initial structure of a cloud may affect its dynamics and lifetime. \cite{SR17} (see also \citealt{SR15}) examined that a turbulent cloud is destroyed more quickly than homogeneous cloud in the hot wind. We adopted a random density perturbation $\delta \rho/\rho$ distributed between $-0.25$ and 0.25 for all the simulations in Section \ref{section_cloudmodels}. Here we carry out a simulation for a turbulent cloud. Using the initial and boundary conditions in T1L, in the cloud we apply an initial perturbed turbulent velocity field following a Gaussian random distribution with a Fourier power $|v^{2}(k)|\propto  k^4$ (e.g., \citealt{Ostriker01}; \citealt{Gong11}; \citealt{Chen15}). The initial velocity turbulence may cause density turbulence in the cloud, meanwhile the cloud is accelerated by radiation. The filamentary structure formed during cloud acceleration in this run is twisted due to the initial turbulence, thus the morphology of the cloud changes by cloud turbulence. On the other hand, we find that the cloud survival time $t_{50}\sim 7.7\,t_*$, which is slightly shorter than it for T1L. However, the cloud $t_{50}$ is still significantly longer than that in the hot wind. Also we find that the change of $V_{\rm mean}$ due to turbulence is about $\sim 4\%$ at $t=7\,t_*$. Therefore we conclude that the effects of turbulence on cloud acceleration and survival time is insignificant.

Also we assumed the molecular weight $\mu=1$ for our simulations in Section \ref{section_results}. However, molecular weight may vary for different state of gas. For example, molecular weight $\mu \simeq 0.6$ for fully ionized gas with a solar-metallicity, while $\mu \simeq 2.33$ for neutral molecular gas. Here we perform another run based on T1L but with $\mu = 2.33$ in the molecular cloud, and $\mu =0.6$ in the fully ionized hot background, thus the background temperature has a lower value o $T_{\rm bkgd}= (0.6/2.33)\chi_0 T_* \simeq2.6\times10^{5}\,$K. We find a slightly lower acceleration in the run with changed $\mu$ with $V_{\rm mean}$ decreasing to $\sim 95\%$ and $\sim 90\%$ of its values at $t=7\,t_*$ and $t=8\,t_*$ in T1L respectively. Also we find that the cloud mass evolution $M_{c}/M_{c,\rm i}$ shows very similar behavior in this run compared to T1L, suggesting that changing $\mu$ does not change our conclusions. 

\begin{figure}[t]
\centerline{\includegraphics[width=9cm]{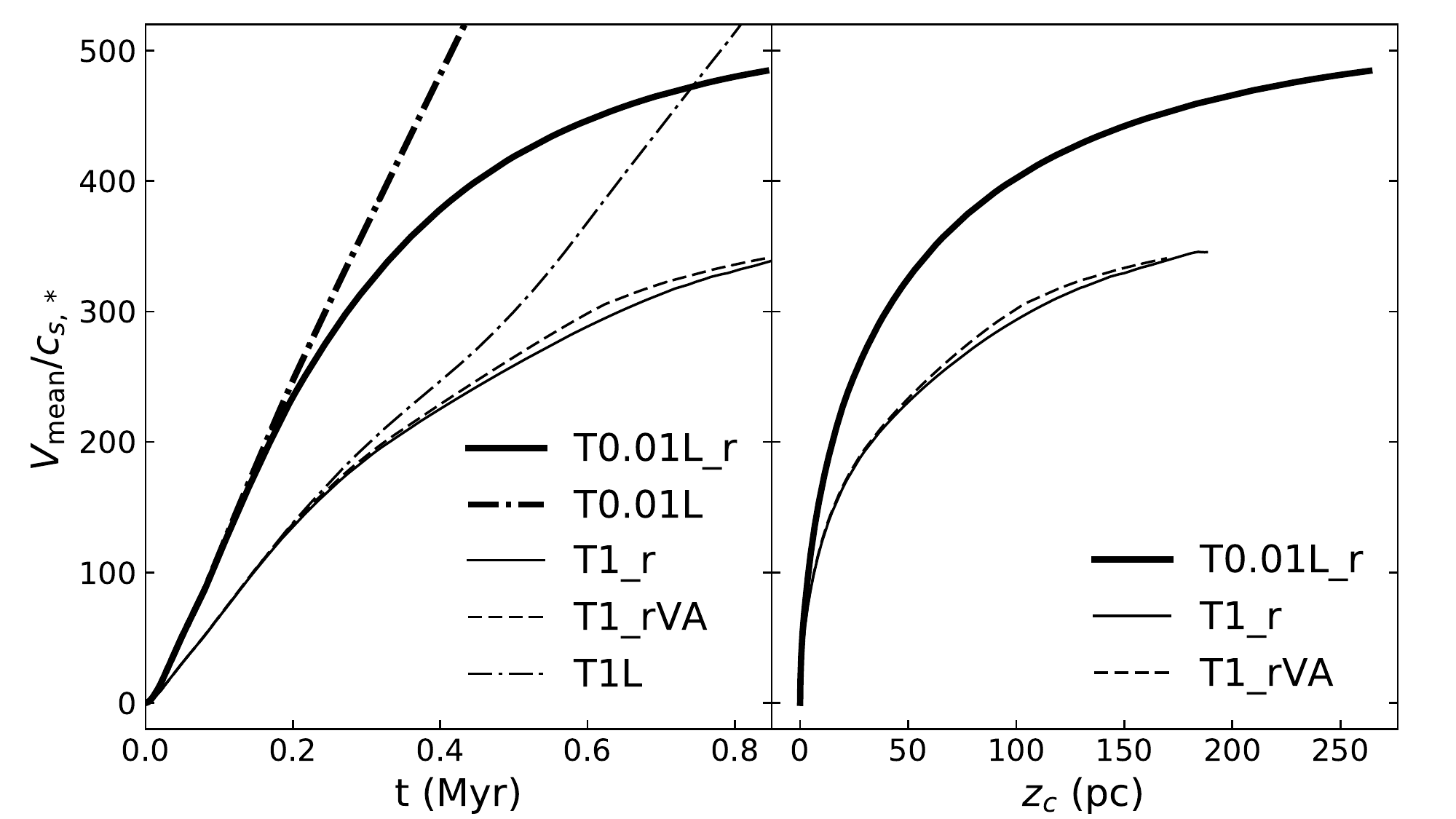}}
\caption{Cloud mean velocity as a function of time (left panel) and flying distance $z_{c}$ (right panel) for runs T0.01L\_r, T1L\_r with hot background and T1L\_rVA with vacuum background, where the flux is given by equation (\ref{fluxvar}) and $R=200\,$pc. The velocity $V_{\rm mean}$ from T0.01L and T1L also plotted in the left panel.}\label{fig_sphere}
\end{figure}

\subsection{Flux Variation and Vacuum Background}\label{section_secondfactors2}

If the cloud flying distance $z_{c}$ (see Section \ref{section_initial}) is comparable to the size of the radiation-dominated region of the galaxy, the assumption of a constant flux boundary breaks down. For example, \cite{Thompson15} studied the acceleration of dusty radiation-pressure-driven clouds with a spherical flux $F \propto r^{-2}$ semi-analytically, where clouds are assumed as spheres, and clouds expand in vacuum freely with its sound speed. Here, we perform four runs (T0.01L\_r, T0.01L\_rVA, T1L\_r and T1L\_rVA) which have the same setup as T0.01L and T1L respectively, but the boundary flux injection satisfies
\begin{equation}
F=F_*\frac{R^{2}}{(R+z_c)^{2}}\label{fluxvar},
\end{equation}
where we choose the size of the starburst region as $R = 200\,$pc. Also we consider two types of background medium, one is the hot background ($\chi_0=10^{4}$ for T0.01L\_r and T1L\_r), another is the ``vacuum background" (T0.01L\_rVA and T1L\_rVA), in which we set $T_{\rm bkgd}=T_*$ and a low density $\rho = 10^{-6}\rho_*$ in the background. The pressure of the cloud cannot be confined by the vacuum background so it can expand freely with its internal sound speed. Figure \ref{fig_sphere} shows the results of $V_{\rm mean}$ as a function of time and flying distance. We stop the simulations at $t=8\,t_*$ but we use Myr and pc instead of code time and length units for Figure \ref{fig_sphere}. T0.01L\_r and T0.01L\_rVA show almost identical acceleration history so we only plot one of them. The difference between $V_{\rm mean}$ from T1L\_r and T1L\_rVA is also negligible, which means that free expansion does not change the dynamics of cloud up to $\tau_*=1$.  The cloud velocities are significantly lower than $V_{\rm mean}$ from constant-flux runs. Note that since $F_*$ decreases along the vertical direction, the dust temperature also drops. We find that $T_{c}\sim (F_*/4)^{1/4} T_c \sim 0.85 T_*$ at $z_c = R$, thus dust in the cloud is still in thermal equilibrium with the radiation field. The decreases opacity combines with dropping flux together to affect the cloud velocity. Cloud with $\tau_* =0.01$ can be accelerated to $z_{c}\simeq R= 200\,$pc with $V_{\rm mean} \sim 470\,c_{s,*}$ (420\,km s$^{-1}$) at $t\sim 0.7\,$Myr, while clouds with $\tau_* = 1$ can be accelerated to $z_{c}\simeq R=200\,$pc at $\sim 0.9\,$Mpc with a velocity of $V_{\rm mean} \sim 350\,c_{s,*}$ (310\,km s$^{-1}$). Compared to Figures \ref{fig_T001L} and \ref{fig_T1L}, the cloud velocity $V_{\rm mean}$ drops to $\sim 60\%$ of its value at the same radial position compared to the constant flux run. We also check that for a larger star-forming core $R=500\,$pc $V_{\rm mean}$ decreases to $\sim 75\,\%$ of its value for constant-flux case at $z_{c}= R$.

\begin{figure}[t]
\centerline{\includegraphics[width=8.5cm]{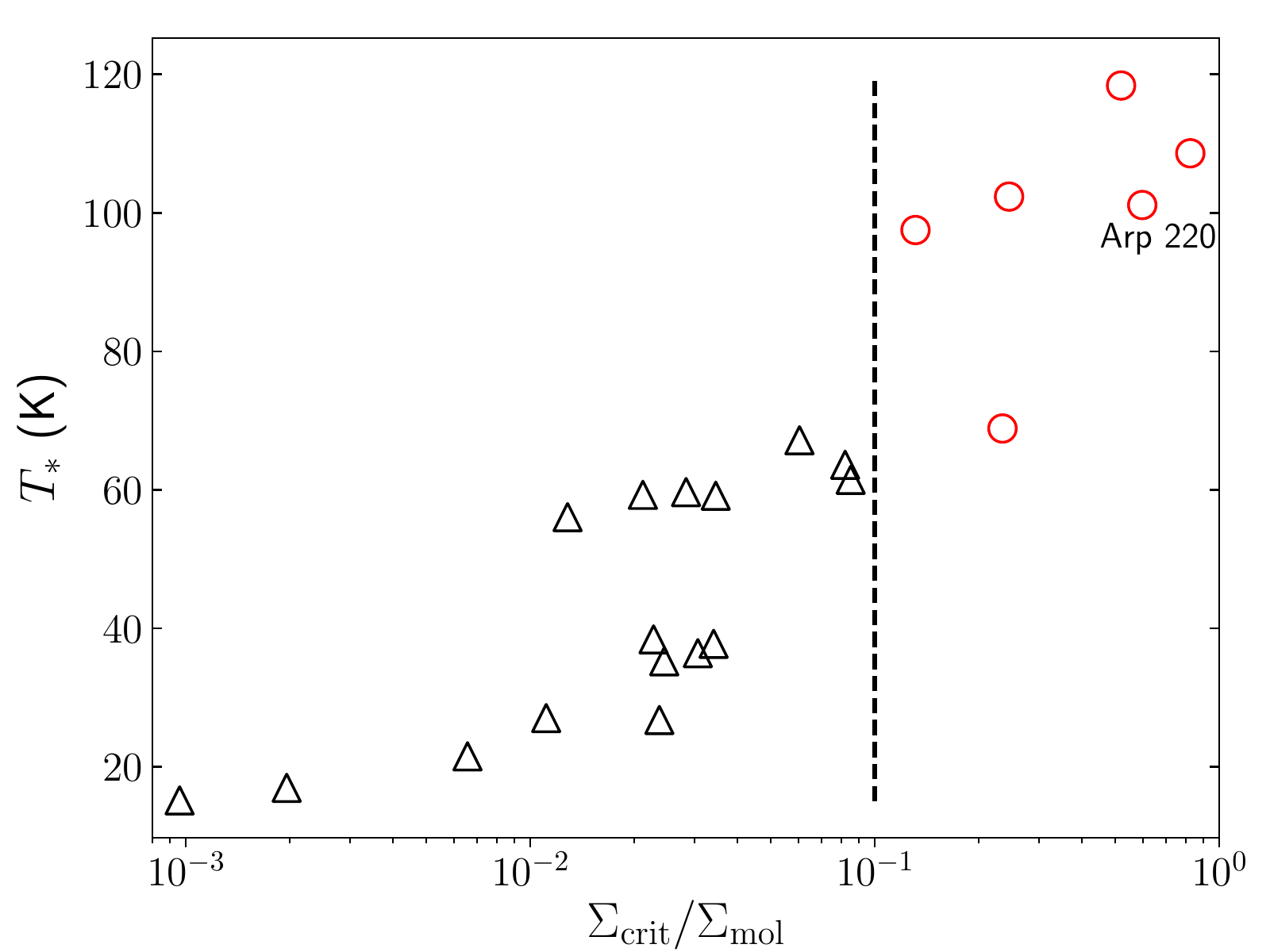}}
\caption{Estimated dust temperature vs. Eddington ratio $\Sigma_{\rm crit}/\Sigma_{\rm mol}$ using samples from \cite{Barcos17}. Triangles are data with $\Sigma_{\rm crit}/\Sigma_{\rm mol}<0.1$ and cycles are data with $\Sigma_{\rm crit}/\Sigma_{\rm mol}>0.1$. Also the starburst galaxy Arp 220 is marked.}\label{fig_data}
\end{figure}

\subsection{Implications to LIRGs and ULIRGs}\label{section_ULIRGs}

The pancake structure of clouds along the line of sight has been observed via narrow absorption line (NAL) or broad absorption line (BAL) in AGN environments (e.g., \citealt{Hall07}; \citealt{Rogerson11}; \citealt{Hamann13}). \cite{Proga14} performed radiation hydrodynamics to model the pancake structure of the irradiated clouds in AGNs. Analogously, we suggest that future observations of cloud morphologies may give a hint on the mechanism of cloud acceleration in LIRGs and ULIRGs. Optically thin clouds with column density $N_{\rm H}\sim 10^{21}\,$cm$^{-2}$ has the typical columns observed in galactic winds (\citealt{Martin05, Martin15}). We find that these clouds develop pancake structure due to radiation pressure on dust, while the entrained clouds in hot flows fragment into small pieces and stretch out along the direction of motion. The different theoretical predictions of cloud evolution can be potentially used to guide future observations.   

The gravity contributed by galactic disk, bulge and halo can further decreases $V_{\rm mean}$. Dust in a cloud can be considered to be in thermal equilibrium with the radiation field, therefore the infrared flux from the galactic disk can be calculated by equation (\ref{dimensionless1}). The critical disk density can be given by $2\pi G \Sigma_{\rm crit}=g_*$, where $g_*$ is from equation (\ref{dimensionless2}). Combining equations (\ref{dimensionless1}) and (\ref{dimensionless2}) we obtain
\begin{equation}
\Sigma_{\rm crit}= 2.7\times 10^{4}T_{*,2}^{6}\,M_{\odot}\,\textrm{pc}^{-2}.\label{Sigmacrit}
\end{equation} 
If the galactic surface density $\Sigma > \Sigma_{\rm crit}$, the LIRG/ULIRG system is a sub-Eddington system. Note that the opacity we used (equation \ref{opacity}) assumes a Milk-Way-like dust-to-gas ratio. The actually $\Sigma_{\rm crit}$ is proportional to the dust-to-gas ratio, which is still uncertain in most galaxies. 

We have used a fiducal dust temperature $T_*=100\,$K in all simulations. In fact, the cloud temperature can be estimated from the observed flux, assuming thermal equilibrium between radiation and dust in the atmosphere of the galaxy. Taking data from \cite{Barcos17} (Table 6) as samples, we estimate that the temperature of dusty clouds has a range of $T_c\sim 15-120\,$K. Also we can calculate $\Sigma_{\rm crit}$ using equation (\ref{Sigmacrit}), and compared with molecular surface density $\Sigma_{\rm mol}$ which gives the lower bound of galactic surface density $\Sigma$. As mentioned in \cite{ZD17}, we find all the samples in \cite{Barcos17} give $\Sigma_{\rm crit}<\Sigma_{\rm mol}$, thus all of them are sub-Eddington systems. However, \cite{Thompson16b} showed that a fraction of gas in the star-forming core can still be locally super-Eddington even for a sub-Eddington system. The mass probability distribution for $\Sigma_{\rm crit}/\Sigma >0.1$ is $\sim 10^{-3}-2\times 10^{-2}$, while the area probability distribution for $\Sigma_{\rm crit}/\Sigma >0.1$ is $\sim 0.04-0.4$, depending on the property of gas turbulence. This scenario can explain the origin of cloud launching by radiation pressure from a sub-Eddington system. Figure \ref{fig_data} shows the estimated $T_*$ vs. global Eddington ratio $\Sigma_{\rm crit}/\Sigma_{\rm mol}$ for the samples in \cite{Barcos17}. Interestingly, we find that the dust temperature correlates with the Eddington ratio $\Sigma_{\rm crit}/\Sigma_{\rm mol}$, and most galaxies with $\Sigma_{\rm crit}/\Sigma_{\rm mol}>1$ have $T_* \sim 100-120\,$K. For example, we estimate that Arp 220 has $\Sigma_{\rm crit}/\Sigma_{\rm mol}\sim 0.6$ with $T_{c}\sim 100\,$K, a value between the warm dust temperature in eastern and  western nuclei (e.g., \citealt{Wilson14}; \citealt{Scoville15}). We estimate the upper bound velocity for a 10-pc size cloud with a decreasing factor $\epsilon$ compared to the case of constant flux with gravity can reach
\begin{eqnarray}
V_{c}\sim 310\,\textrm{km s}^{-1}&&\left(\frac{\epsilon}{0.6}\right)\left(\frac{t}{0.7\,\textrm{Myr}}\right)\nonumber\\
&&\times \left(\frac{F_*}{5.6\times 10^{13}\,L_{\odot}\textrm{kpc}^{-2}}\right)^{3/2},
\end{eqnarray}
which may explain the observations of molecular outflows in LIRGs and ULRIGs (\citealt{Sakamoto99}; \citealt{Walter02}; \citealt{Veilleux05} and references therein; \citealt{Veilleux09}; \citealt{Fischer10}; \citealt{Bolatto13}; \citealt{Cicone14}; \citealt{Walter17}).

\section{Conclusions}\label{section_conclusions}

We study cold clouds accelerated by ram pressure on dust in the environment of rapidly star-forming galaxies dominated by infrared radiation flux. We perform a series of 2D and 3D radiation hydrodynamic simulations, utilizing the reduced speed of light approximation to solve the frequency-averaged, time-dependent radiative transfer equation. The radiative acceleration of a cloud in pressure equilibrium with a surrounding hot, tenuous medium can be described by 
\begin{equation}
V_{\rm mean} = \epsilon \frac{\kappa_{R} F_*}{c}t,
\end{equation}
where $\kappa_R$ is the Rosseland mean opacity for the dust, $F_*$ is the radiation flux (equations \ref{analytic} and \ref{analytic2}). The coefficient $\epsilon$ mainly depends on the initial optical depth of the cloud. We obtain $\epsilon\sim 1.4$ for very optically thin clouds, and $\epsilon$ decreases from $\epsilon \sim 0.8$ to $\epsilon \sim 0.18$ for optically thick clouds from $\tau_* = 1$ to $\tau_* = 10$. Empirically we have $\epsilon \simeq \textrm{min}\{1,\exp(1.3-1.5\tau_*^{0.3})\}$ to fit cloud acceleration from optically thin to thick clouds up to $\tau_* \sim 10$ (equation \ref{taufitting2}). However,the gas pressure in the interface between the background and the front of the cloud may decelerate and destroy the irradiated cloud if the cloud is embedded in a warm medium $T_{\rm bkgd}\lesssim 10^{4}\,$K that the initial density ratio between the cloud and background $\chi_0 = \rho_*/\rho_{\rm bkgd}\lesssim 10^2$ (Figure \ref{fig_warmISM}).

The evolution of the cloud geometry and morphology during its acceleration depends on its characteristic lengthscale and other attributes. In general, an optically thin cloud with a size of $\sim10\,$pc is squeezed in the direction of motion to form a pancake structure (Figure \ref{fig_T001L}), which is caused by the combination of radiation pressure, gas pressure from the interface between the front of the cloud and the background, as well as ram pressure of the background. On the other hand, differential acceleration of an optically thick, $\sim 10$ pc cloud forms a filamentary shape elongated parallel to the acceleration direction extending to $\sim 100\,$pc (Figure \ref{fig_T1L}). A much smaller cloud with the same initial column density/optical depth and same $\chi_0$, but smaller characteristic lengthscale may also form a pancake or filamentary structure, but eventually fragments into small pieces due to shearing instabilities between the cloud and background (Figure \ref{fig_SR}). The morphology of cloud is somewhat sensitive to various assumptions: different reduction factors of the reduced speed of light, spatial resolution, cooling and heating processes in the background, initial turbulence profile, and perturbation inside the cloud, as well as the dimensionality of the simulation. However, the dynamics of cloud acceleration is only weakly affected by these assumptions, unless a dense background $\chi_0 \lesssim 10^{2}$ is present to shred the cloud. 

We also compare the dynamics of the cloud accelerated by radiation fields to an entrained cloud in a hot flow driven by supernovae in the host galaxy in a limit where the momentum injection of the hot flow is the same as the momentum of radiation. The cloud survival time is defined as half of the initial cloud mass is still above 1/3 of its initial average density. Although the survival time of clouds in hot flows is shorter in our hydrodynamic simulations compared to the literature, and the cloud is shredded much faster in a warm medium compared to a hot medium, we find the the survival time is still significantly longer for irradiated cloud than the hot-flow-entrained cloud even including all the uncertainties (Figures \ref{fig_mass} and \ref{fig_hotcloud}). Therefore, a cloud in a radiation-dominated region can be accelerated to a higher velocity with a larger distance compared to a cloud in the hot flow environment. This result can be apply to LIRGs and ULIRGs, and we find that dusty clouds in LIRGs and ULRIGs can be accelerated to hundreds of km s$^{-1}$ and potentially match the observations.

\acknowledgments

We thank Todd Thompson, Daniel Proga, Peng Oh, Mike McCourt, Sylvain Veilleux, Eliot Quataert, Norm Murray, Tim Waters, Suoqing Ji, Patrick Hall, Ari Laor and Sebastian Hoenig for stimulating discussions. This work used the computational resources provided by the Advanced Research Computing Services (ARCS) at the University of Virginia. We also used the Extreme Science and Engineering Discovery Environment (XSEDE), which is supported by National Science Foundation (NSF) grant No. ACI-1053575. S. W. D. and D. Z. acknowledge support from NSF grant AST-1616171 and an Alfred P. Sloan Research Fellowship. Y. F. J is supported part by the National Science Foundation under Grant No.~NSF PHY 17-48958. J. M. S. is supported by NSF grant AST-1333091.

\appendix

\section{Reduced Speed of Light Approximation}\label{append_reduced_speed}

The dimensionless radiation hydrodynamics equations include two ratios $\Crat=c/a_*$ (see the definition in Section \ref{section_reduced_speed1}) and $\Prat = a_r T_*^{4}/P_*$, where $a_*$, $T_*$ and $P_*$ are the characteristic velocity, temperature and pressure respectively (\citealt{Jiang12}). The dimensionless time-step $\Delta t$ in the explicit algorithm is constrained by the Courant condition
\begin{equation}
\Delta t < C_{\rm CFL}\frac{\Delta z}{\Crat}\label{courant},
\end{equation}
where the dimensionless $\Delta z$ is the cell width in the computational box (see Table \ref{tab_parameter}). In the paper we fix the Courant--Friedrichs--Lewy number to be $C_{\rm CFL}=0.4$ for 2D runs expect for T1LR3$_{\rm lowCFL}$, and $C_{\rm CFL}=0.3$ for 3D runs (see also \citealt{Jiang14}). Therefore, for cloud $T_* = 100\,$K and $a_0=c_{s,*}$,  we have $\Crat \simeq 3.3\times 10^{5}$, and the Courant condition equation (\ref{courant}) gives a severe constraint on $\Delta t$ that $\Delta t < 8\times10^{-7}\Delta z$. In order to reduced the computational costs, we use the reduced speed of light approximation, which has been implemented in radiation hydrodynamical simulations (\citealt{Gnedin01}; \citealt{Gonzales07}; \citealt{Aubert08}; \citealt{Petkova11}; \citealt{Rosdahl13}; \citealt{Skinner13, Skinner15}; \citealt{Gnedin16}). The parameter $\tilde\Crat$ is used instead of $\Crat$ before the radiation source terms (see equations \ref{reduced_radsource1} in Section \ref{section_reduced_speed1}).

The domain of validity for the reduced speed of light approximation is well described in Section 3.2 of \cite{Skinner13} along with references therein. The main logic of the approximation is that signal propagation on the light crossing time is often irrelevant in systems where the radiation transfer is dominated by local interactions between the radiation field and matter.  The key constraint in these systems is then on maintaining the correct ordering of dynamic timescale $t_{\rm dyn} \sim L/v_{\rm max}$ and the diffusion timescale $t_{\rm diff} \sim L \tau_{\rm max}/c$, where $L$ is the characteristic size of the system. To maintain the correct ordering, the lower bound of $\tilde\Crat$ is constrained by
\begin{equation}
\tilde\Crat \gg \frac{v_{\rm max}}{a_*}\textrm{max}\{1,\tau_{\rm max}\}\label{skinner}.
\end{equation}
The main difficulty with enforcing this criterion is determining the appropriate values for $v_{\rm max}$ and $\tau_{\rm max}$, which are nominally the maximum velocity and optical depth in the system, respectively.  In practice, $\tau_{\rm max}$ requires a choice of a characteristic length over which to define the optical depth.  A conservative approach would take the the maximum optical depth anywhere in the system.  However, this may still result in a timestep that is unnecessarily small if the largest velocities occur only in optically thin regions.  Therefore, the choice for the reduction factor requires careful consideration and testing.

In order to justify the reduced speed of light approximation and confirm the validity of the above criterion, we use the radiation \textsc{athena++} code to do several test simulations. We first consider the impact on radiative linear waves, where we can analytically compute the dispersion relation in the reduced speed of light approximation and then consider the more complex case of dusty gas accelerated by radiation pressure. 

\begin{figure}[t]
\centerline{\includegraphics[width=9cm]{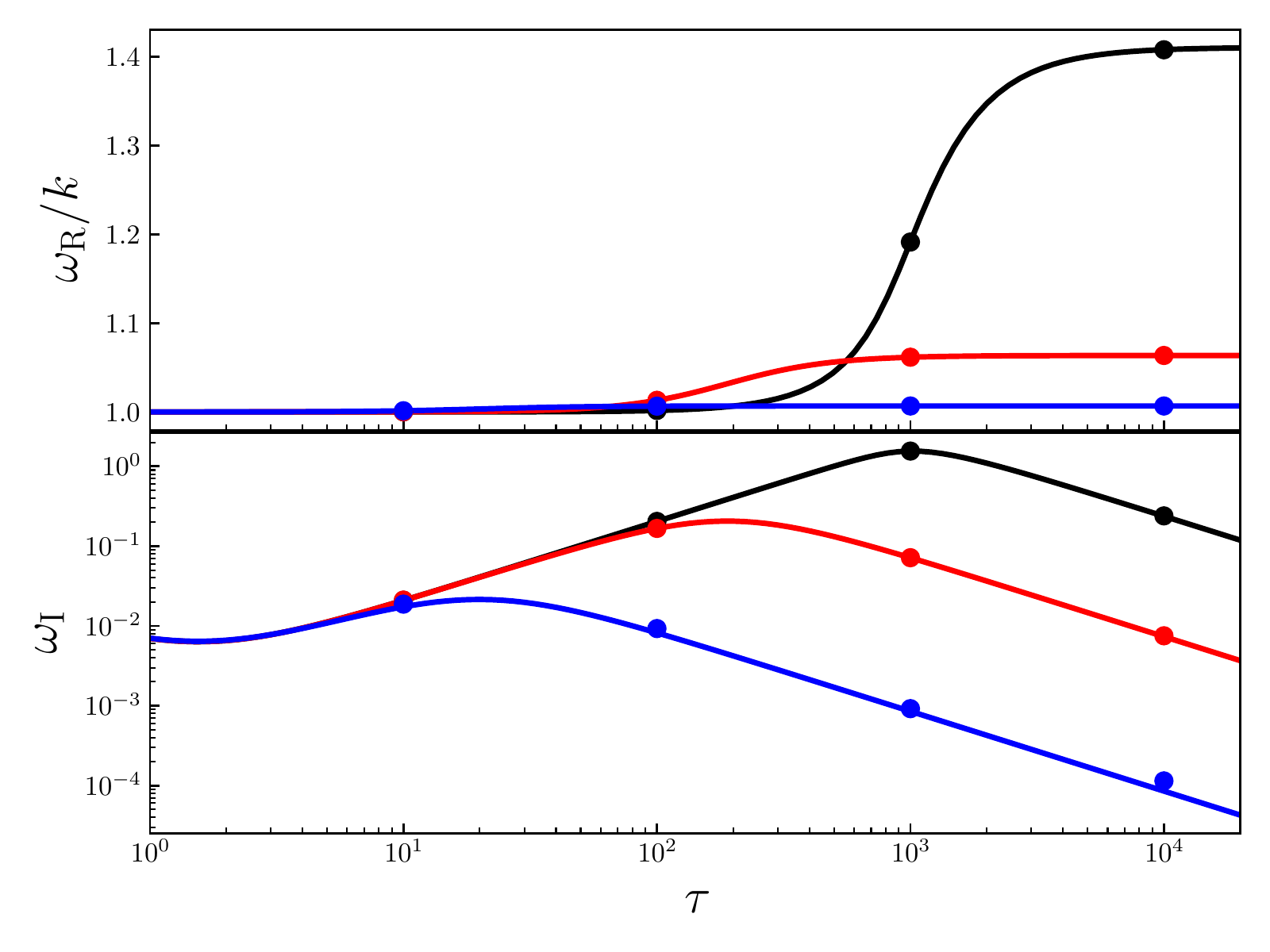}}
\caption{Comparison of the wave propagation speed (upper panel) and the damping rate of linear waves (lower panel) as function of optical depth per wavelength, where $\Prat =1$, $\Crat = 1000$, and the reduction factor $\Rrat =1$ (black lines), 0.1 (red lines) and 0.01 (blue lines). The curves are the analytical solutions, and the circles are simulation results given by \textsc{athena++}. Note that our resolution study implies that the slight mismatch between the simulation and analytic curve for $\Rrat=0.01$ at $\tau=10^4$ is due to the limits of resolution giving rise to numerical diffusion and does not imply a problem with our implementation of the reduced speed of light approximation.}\label{fig_linear}
\end{figure}

\subsection{Linear Wave Test}

We test the 1D linear wave in radiation hydrodynamics. The dispersion relation between $\omega$ and $k$ is given by \cite{Johnson10} (see also \citealt{Jiang12}), where $\omega$ and $k$ are the angular frequency and the wavenumber respectively. We generalize these derivations in a straightforward manner to include the reduction factor $\Rrat$.  The background is setup to be $\rho = P= T =1$, the ratios $\Prat$ and $\Crat$ were varied, but for brevity we only show the case with $\Prat=1$ and $\Crat = 1000$. We choose the reduction factor $\Rrat = 1$, 0.1 and 0.01, which gives $\tilde\Crat = 1000, 100$ and 10 respectively. Figure \ref{fig_linear} shows the effects of reduced speed of light depending on the optical depth per wavelength in the linear wave system. Since $\omega$ can be a complex number, the ratio of its real part and $k$ that $\omega_{\rm R}/k$ defines the wave propagation speed, and the imaginary part $\omega_{\rm I}$ defines the wave damping rate. Note that the curves in Figure \ref{fig_linear} are given by analytic results, and the circles represent the values of $\omega_{\rm R}$ and $\omega_{\rm I}$ that are obtained by fitting the \textsc{athena++} results after running for one adiabatic wave period.  We see the effects of the reduced speed of light approximation in both the phase velocity and the damping rate, with the most significant effect in the damping rates. The damping rate should turn over at $\tau\sim 10^{3}$ with $\Crat = 1000$, but it turns over at lower optical depth at $\tau \sim \tilde\Crat$ for reduced speed of light cases. We considered other values of $\Crat$ (not shown) and confirm that this results holds quite generally.  Hence, the reduced speed of light approximation is valid for $\tilde\Crat<\tau$. Since $v_{\rm max}\simeq a_*$, this result is consistent with equation (\ref{skinner}).  We find that the \textsc{athena++} results agree quite well with the curves, and confirm convergence in the L1 norm as resolution increases.  These results would tend to confirm that the reduced speed of light approximation is implemented correctly in the \textsc{athena++} radiation module.

\begin{figure}[t]
\centerline{\includegraphics[width=9cm]{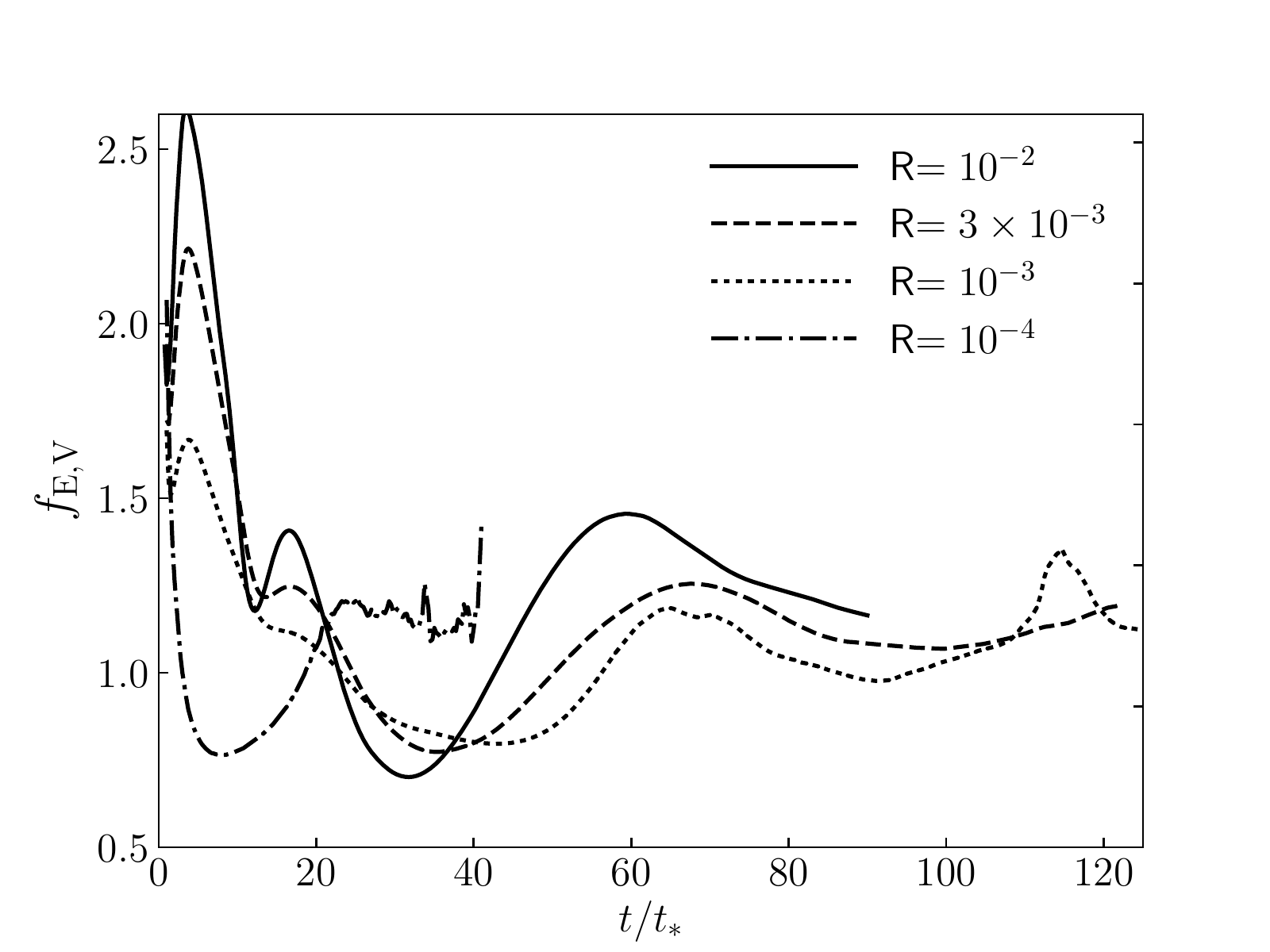}}
\caption{Volume-weighted Eddington ratio $f_{\rm E,V}$ as a function of time for four runs $\Rrat=10^{-2},10^{-3},3\times 10^{-3}$ and $10^{-4}$. The initial and boundary conditions are the same as in the T3\_F0.5 run in \cite{Davis14}.}\label{fig_kt}
\end{figure}

\subsection{Radiation-Pressure-Driven Galactic Winds}

Radiation pressure on dust can drive large-scale galactic winds (\citealt{Murray05}). In particular, the problem of the two-dimensional radiation hydrodynamic of a column of gas that is accelerated by a constant infrared radiation flux has been studied first by \cite{KT12,KT13}, then by others using a variety of algorithms (\citealt{Davis14}; \citealt{Tsang15}; \citealt{Rosdahl15}; \citealt{ZD17}). Here we redo this problem using the radiation \textsc{athena++} code. The initial and boundary conditions are the same as the T3\_F0.5 run in \cite{Davis14}: we assume the initial infrared optical depth of the dusty gas is $\tau_*=3$, the gas temperature is $T_*=82\,$K, the infrared flux $F_*=a_r c T_*^{4}$, the dusty opacity $\kappa_{\rm R}$ follows equation (\ref{opacity}) in this paper, and the initial dimensionless Eddington ratio is setup as $f_{\rm E,*}=\kappa_{\rm R} F_* /(gc)$, where $g$ is the gravity. The scale height is $h_*=c_{s,*}^{2}/g$, and the size of the box is $[L_x \times L_y]/h_*=500\times3200$ with a resolution of $\Delta x/h_*=1$. We choose the reduced speed of light $\Rrat=10^{-2},3\times 10^{-3},10^{-3}$ and $10^{-4}$, and carry out four runs. The runs stops once the gas hits the top of the box expect for the run with $\Rrat=10^{-4}$. We find that the gas eventually becomes unbound. This result is similar as that in \cite{Davis14}. An important parameter to measure the properties of a unbound gas is the Eddington ratio
\begin{equation}
f_{\rm E, V}=\frac{\langle \kappa_{\rm R} \rho F_ \rangle}{c g \langle \rho \rangle}.
\end{equation}
Figure \ref{fig_kt} shows $f_{\rm E,V}$ as a function of time. We find although there are initial bumps at early time, the Eddington ratio $f_{\rm E,V}$ converges to $f_{\rm E,V}\sim 1$ for $\Rrat \geq 10^{-3}$ at later time. The reduced speed of light approximation breaks down at $\Rrat=10^{-4}$. We stops $\Rrat=10^{-4}$ run at $t\simeq 40\,t_*$, otherwise the solution becomes unstable. We use the vertically integrated optical depth to estimate $\tau_{\rm max}$ and find that $\tau_{\rm max}\sim 20$. Also, we get $v_{\rm max}/c_{s,*}\sim 20$, thus the criterion equation (\ref{skinner}) gives $\tilde\Crat \gg 400$, or $\Rrat \gg 7\times 10^{-4}$. This result is consistent with Figure \ref{fig_kt}.


\clearpage

\end{document}